\documentclass[preprint2]{aastex}
\usepackage{times}
\usepackage{amsmath}
\usepackage{xcolor}
\usepackage[hidelinks]{hyperref}

\usepackage{chngcntr}

\newcommand{\me}{$M_{\oplus}$}

\newcommand{\teff}{$T_{\rm eff}$}

\newcommand{\teq}{$T_{\rm eq}$}

\newcommand{\tint}{$T_{\rm int}$}

\newcommand{\cas}{\emph{Cassini}}
\newcommand{\juno}{\emph{Juno}}

\definecolor{grau}{rgb}{0.6,0.6,0.6}
\definecolor{gruen}{rgb}{0,0.8,0}

\definecolor{orange}{HTML}{FFA500}

\newenvironment{packed_enum}{
\begin{itemize} 
  \setlength{\itemsep}{1pt}
  \setlength{\parskip}{0pt}
  \setlength{\parsep}{0pt}
}{\end{itemize}}

\begin{document}

\counterwithin{figure}{section}
\counterwithin{table}{section}
\numberwithin{equation}{section}

\setcounter{section}{4}

\title{\textbf{\LARGE 4}}
\title{\textbf{\LARGE  }}
\title{\textbf{\LARGE Saturn's Interior After the Cassini Grand Finale}}

\author{J. J. Fortney, B. Militzer, C. R. Mankovich, R. Helled, S. M. Wahl, N. Nettelmann, W. B. Hubbard, D. J. Stevenson, L. Iess, M. S. Marley, N. Movshovitz }










\begin{abstract}
\baselineskip = 11pt
\leftskip = 0.65in 
\rightskip = 0.65in 
\parindent=1pc
\end{abstract}  

{\textbf{\normalsize Abstract}} \\We present a review of Saturn's interior structure and thermal evolution, with a particular focus on work in the past $\sim$5 years.  Data from the \cas\ mission, including a precise determination of the gravity field from the Grand Finale orbits, and the still ongoing identification of ring wave features in Saturn's C-ring tied to seismic modes in the planet, have led to dramatic advances in our understanding of Saturn's structure.  Models that match the gravity field suggest that differential rotation, as seen in the visible atmosphere, extends down to at least a depth of 10,000 km (1/6$^{\rm th}$ the planet's radius).  At greater depths, a variety of different investigations all now point to a deep Saturn rotation rate of 10 hours and 33 minutes.  There is very compelling evidence for a central heavy element concentration (``core''), that in most recent models is 12-20 Earth masses.  Ring seismology strongly suggests that the core is not entirely compact, but is dilute (mixed in with the overlying H/He), and has a substantial radial extent, perhaps out to around one-half of the planet's radius.  A wide range of thermal evolution scenarios can match the planet's current  luminosity, with progress on better quantifying the helium rain scenario hampered by Saturn's poorly known atmospheric helium abundance. We discuss the relevance of magnetic field data on understanding the planet's current interior structure.  We point towards additional future work that combines seismology and gravity within a framework that includes differential rotation, and the utility of a Saturn entry probe. 

\subsection{Introduction} 
Understanding the structure and interior composition of Saturn is a long-standing problem in planetary sciences.  Saturn is one of the solar system's two examples of hydrogen/helium (H/He) dominated planets, so along with Jupiter it serves as a benchmark for our understanding of this class of astrophysical objects.  As such it serves as important point of comparison and contrast with Jupiter, as Saturn formed in the same protoplanetary disk but accreted less gas.  The \emph{Cassini} Mission data obtained and analyzed over the past $\sim$5 years, in particular data from the Grand Finale, have enabled a revolution in our understanding of the planet, even in comparison to the recent pre-Finale review chapter of \citet{Fortney2016}.  The purpose of this contribution is to summarize and discuss this work, from a wide variety of authors, 
while pointing to future work and observational needs.

There are many issues in planetary physics that are addressed by various kinds of interior structure modeling.  We would like to understand:
\begin{packed_enum}
\item The mass of any central heavy element core, along with the current structure of the core itself in relation to the overlying H/He envelope.  This allows for a better understanding of planet formation and the subsequent evolutionary processes \citep{Pollack96,Wilson12b,Helled17}.
\item The distribution of He within the planet.  It has long been suggested that H/He phase separation could have taken place in Saturn leading to composition gradients and He enrichment at depth \citep{SS77b,Hubbard99,FH03,Mankovich20a}.
\item The enhancement and distribution of ``metals" in the H/He envelope, compared to their abundance in the Sun.  The envelope enrichment can be compared to Jupiter \citep{Wong04,Wahl2017a} and  spectroscopic investigation of Saturn's atmosphere to better understand Saturn's formation (chapter by Atreya et al.).
\item The temperature structure of the planet and any deviation from a simple isentropic interior.  Barriers to convection due to abundance gradients \citep{Leconte12,Leconte13,Mankovich20a} can dramatically alter the temperature structure and thermal evolution history \citep{Vazan2018}.
\item The rotation rate of the planet and the depth of any differential rotation.  Rotation uncertainties impact the validity of models and understanding differential rotation helps to understand the mechanics of convection within the planet (chapter by Flasar et al.)
\item The location within Saturn (depth and extent) where the planet's dynamo is found.  A better understanding of the generation of Saturn's magnetic field feeds into better knowledge of the planet's interior structure (chapter by Cao et al.)
\end{packed_enum}

There are several essential inputs for models of the interior structure of giant planets.  These include input physics, meaning the equation of state and mixing properties of hydrogen, helium, and ``heavy elements" like water and other ices, and rocks.  A recent review for hydrogen can be found in \citet{Helled2020}.  A more general overview of the interior modeling of Saturn, including a hierarchy of modeling approaches can be found in \citet{Fortney2016}.  The other needed inputs are observational data that can shed light on the structure and thermal evolution of the planet, and it is on these topics where we will spend time reviewing recent important advances.

\subsection{Observational data during Grand Finale orbits}
\subsubsection{Gravity field observations} \label{sec.grav_obs} 
During the terminal phase of the mission, started in April 2017, \emph{Cassini} was inserted into a series of 22 highly eccentric orbits, with pericenter inside the inner ring of Saturn, at altitudes between 2600 to 3900 km above the cloud tops. From this privileged observation point Cassini carried out the first detailed measurement of the planet's gravity field. 

Due to operational constraints and time sharing with other instruments, only six out of the 22 Grand Finale orbits (GFO) could be used for gravity measurements, which require that radio contact to ground is established via the 4m high gain antenna. In these measurements, the gravity field is inferred from the Doppler shift of a X band microwave signal sent from a ground antenna at 7.2 GHz and coherently re-transmitted back to Earth at 8.4 GHz via an onboard transponder. 
A detailed description of the experimental configuration is described in \citet{Iess2019}. As data from one of the passes were lost due to a station misconfiguration, the actual gravity analysis is based on only five pericenter passes.

The key gravity observable is the Doppler shift of the microwave carrier, which is proportional to the (two-way) range rate of the spacecraft. \emph{Cassini} acts as a probe mass, freely falling in the gravity field of Saturn and the other bodies of its system (most notably, the rings). The radio link allows to observe the line-of-sight accelerations of the spacecraft with respect to a fiducial point (the ground station), from which the gravitational forces acting on the spacecraft can be derived. This well consolidated method to derive planetary gravity fields was used in particular for Jupiter
by the spacecraft \emph{Juno} \citep{Iess2018}.  

Two ingredients are crucial for a successful gravity determination, namely a good dynamical model, accounting for all the forces acting on the spacecraft (described in many cases by free parameters), and high quality Doppler data. The gravity GFO were by design carried out at large solar elongation angles, when charged particle noise from the interplanetary plasma is relatively low. For this reason, the data quality was statistically equivalent in X and Ka band, with a RMS Doppler noise of  0.020-0.088 mm s$^{-1}$ at 30 s integration time, depending on the local conditions at the antenna (mostly wind and water vapour content along the line of sight). This noise is still much smaller (by a factor of 40 and 200) than the Doppler signal produced by weakest the harmonics (J$_3$, J$_{10}$) and Saturn's ring.

\begin{figure}[ht]
	\centering
	\includegraphics[width=0.48\textwidth]{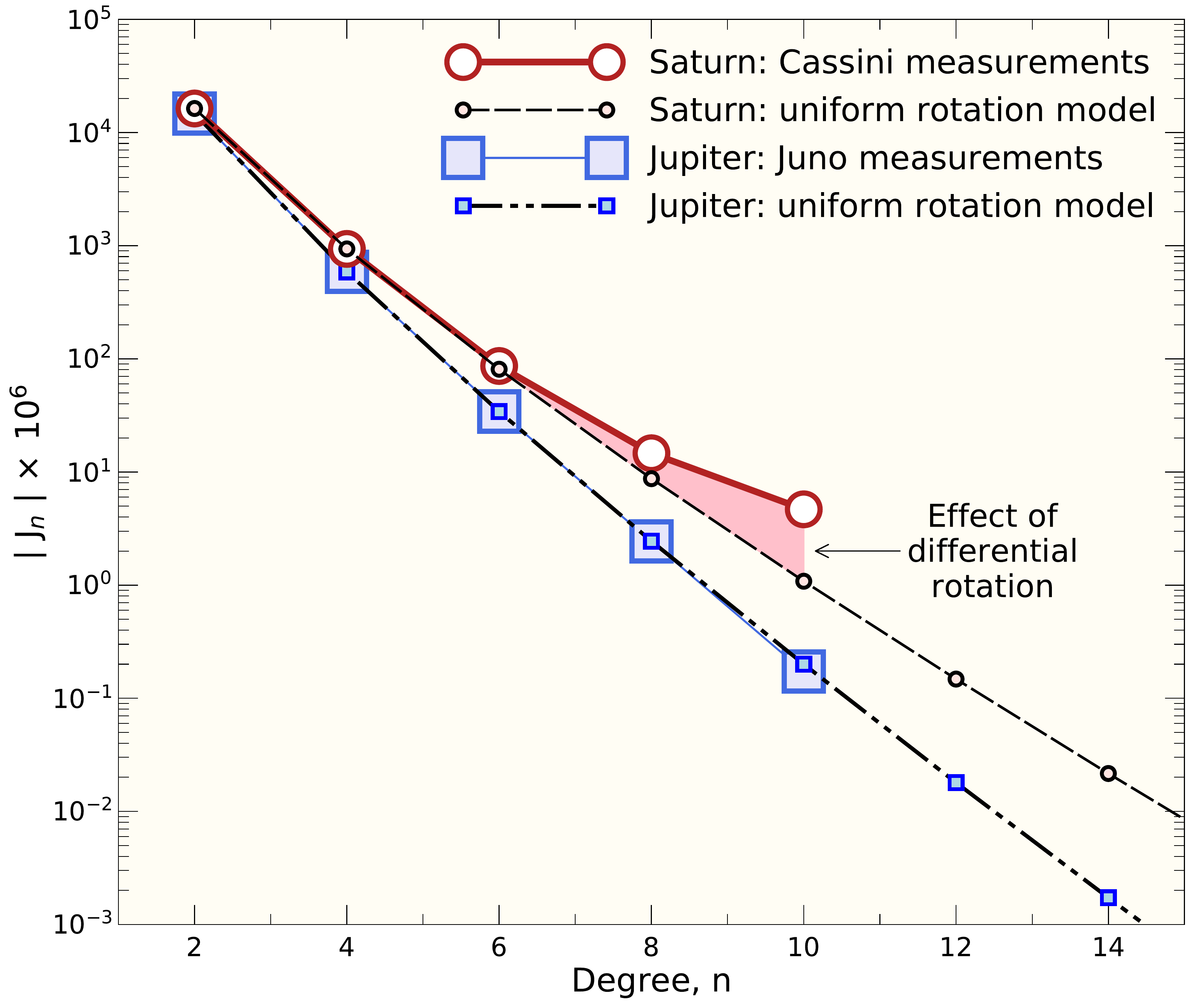}
    \caption{Comparison of the gravity harmonics measured for Jupiter and
    Saturn with predictions from models assuming uniform rotation
  throughout the entire interior of both planets. The deviations are
  small for Jupiter while substantial discrepancies emerge for
  Saturn. This illustrates that the effects of differential rotation
  are much more important for Saturn, as discussed in Section \ref{gravitymod}
      \label{fig.JupiterSaturnJn}.  Models are from \citet{Militzer19}.
    }
\end{figure}

\begin{table*}[ht]
    \begin{center}
\begin{tabular}{| c | r | c | r |} 
\hline
Gravity   & {\it Cassini}~~~~~~~~ & Range of models with      & Model with~~~~~~\\
harmonics & measurements~~~~            & uniform rotation & differential rotation\\
\hline
$J_2$    & 16$\,$290.573 $\pm$ 0.028  & 16$\,$290.57 & 16$\,$290.573~~~~\\
$J_4$    &      $-$935.314 $\pm$ 0.037  &      $-$935.31     &  $-$935.312~~~~ \\
$J_6$    &        86.340 $\pm$ 0.087  & 80.74 $\ldots$ 81.76 & 86.343~~~~ \\
$J_8$    &       $-$14.624 $\pm$ 0.205  & $-$8.96 $\ldots$ $-$8.70 & $-$14.616~~~~\\ 
$J_{10}$ &         4.672 $\pm$ 0.420  &  1.08 $\ldots$  1.13 & 4.677~~~~ \\
\hline
\end{tabular}
    \end{center}
    \caption{The {\it Cassini} gravity measurements are compared with models that assume Saturn's interior to rotate uniformly or differentially. The unexpectedly large magnitudes of $J_6$, $J_8$, and $J_{10}$ could not be matched with uniform rotation models that have a planetary core mass and envelope metallicity adjusted to exactly fit to the observed $J_2$ and $J_4$ ~\citep{Iess2019}.  This is also illustrated in Fig.~\ref{fig.JupiterSaturnJn}.}
    \label{tab:Jn}
\end{table*}

The Doppler data were processed using the JPL MONTE orbit determination code \citep{Evans2016}. The parameters of Saturn's gravity field that were estimated were Saturn's GM, the zonal harmonic coefficients J$_2$-J$_{20}$ (shown in Figure \ref{fig.JupiterSaturnJn}) the tesseral field of degree 2 (allowing for non-principal axis rotation), and the mass of the rings. Although gravity measurements are sensitive to the pole orientation, the pole position was adopted from the more accurate ring occultation measurements of \citet{French2017}. We refer to \citet{Iess2019} for a discussion on the dynamical model and the orbital fit. 

While the purely zonal field absorbs most of the accelerations acting on Cassini 
a surprising finding of the analysis was the impossibility to reduce the Doppler residuals to the noise level in the combined analysis. Residual accelerations of magnitude at the level of $4\times10^{-7}$ m s$^{-2}$ were found within one hour from the epoch of closest approach. By comparison, Juno data could be fit by a purely zonal field, although now, as more passes are accumulated, tiny accelerations about a factor of 10-20 smaller than those experienced by Cassini (2$\times$ $10^{-8}$ m s$^{-2}$) also start showing up in the analysis of Jupiter gravity \citep{Durante20}.  

The nature of these accelerations, a puzzling outcome of the Cassini data analysis, is still unknown. They could be due to gravitational effects caused by longitude-dependent wind dynamics, deep seated gravity anomalies, or normal modes \citep[this list is not exhaustive]{Durante2017,Markham2018,Markham2020} Although normal modes and a high degree tesseral field are able to flatten the Doppler residuals, the preferred method of analysis relied on the use of stochastic accelerations. Remarkably, the estimated values of the zonal field and the B ring mass were nearly insensitive to the method adopted in the analysis.

Understanding the source of these residual fields is one of the open questions for both planets. 
Analysis of oscillations that show up on the surface of the planets --and for Saturn in its rings--can provide unique constraints on internal structure.

\subsubsection{Ring seismological observations}

\begin{figure}[ht]
	\centering
	\includegraphics[width=0.9\columnwidth]{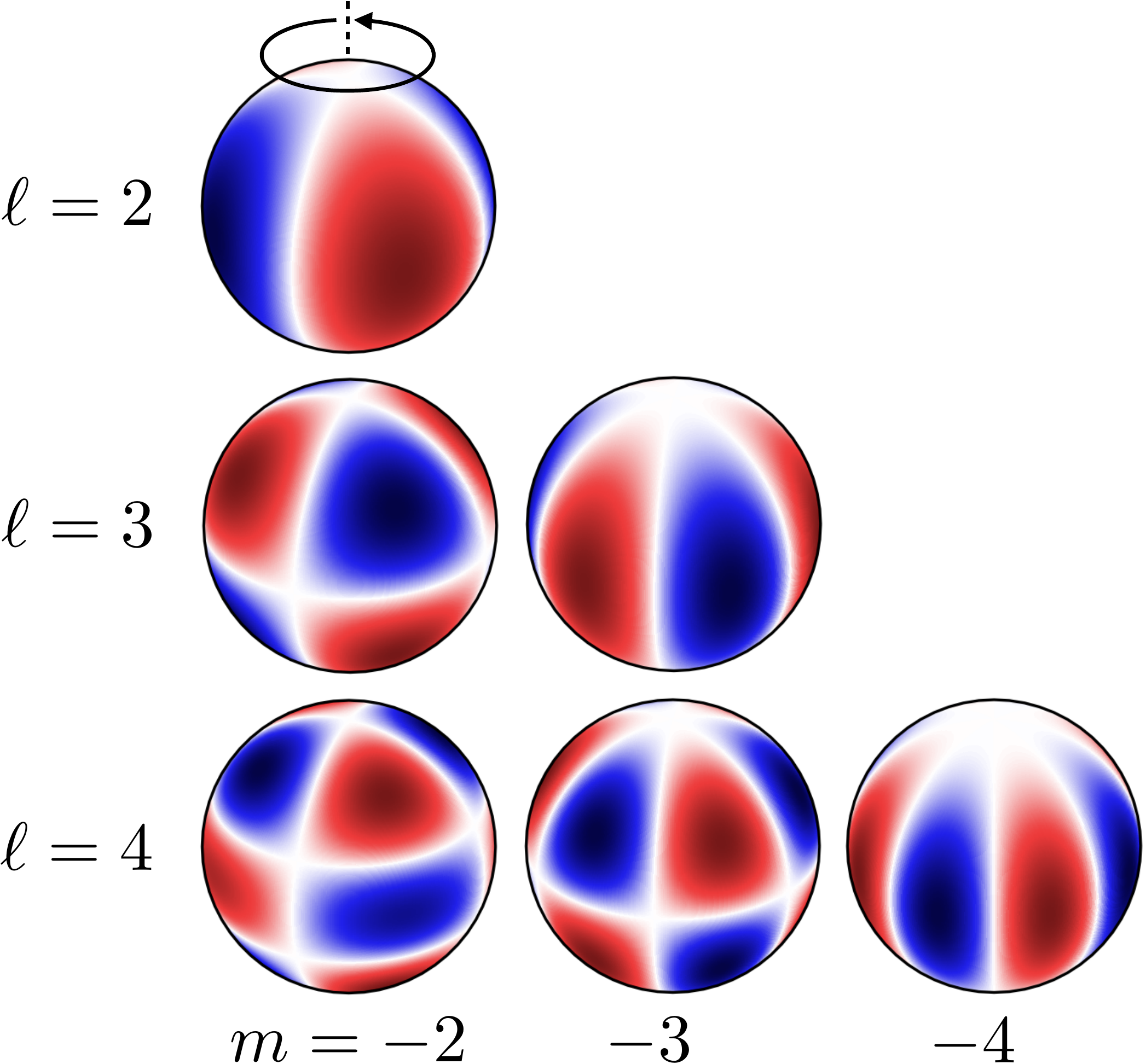}
    \caption{The first few prograde spherical harmonics, a convenient basis for describing the angular structure of Saturn's nonradial modes of oscillation. Angular degree $\ell$ increases downward, and azimuthal order $|m|$ increases to the right. The latitudinal wavenumber $\ell+m$ increases away from the main diagonal $\ell=-m$. Adapted from \cite{Mankovich20b}.
    \label{fig.spherical_harmonics}
    }
\end{figure}

Although \cas's journey ended almost four years before this chapter's writing, ongoing analysis of stellar occultation data gathered by \cas\ VIMS continues to expose ring waves that constrain Saturn's interior.
In their initial study of the potential for Saturn f-modes to excite bending and density waves in the C ring, \citet{Marley93} predicted locations for about a dozen ring features. \citet{Hedman13} confirmed the f-mode -- ring wave connection by associating six waves with planetary oscillation modes. 
Subsequently Hedman, Nicholson, French, and collaborators \citep{Hedman14, French16, French19, Hedman19, French21} have continued to identify ring waves using the entirety of the \cas\ ring data set and now have identified more than thirty waves as being clearly associated with Saturn modes.
In particular these modes are all \emph{non-radial} modes that propagate prograde in Saturn's rotating reference frame, and are conventionally assigned negative azimuthal wavenumbers $m<0$.
The modes span a range of angular degrees $\ell\geq2$, azimuthal wavenumbers $-\ell\leq m<0$, and latitudinal wavenumbers $\ell-|m|=\ell+m$.
Fig.~\ref{fig.spherical_harmonics} provides a visual reference for these labels of angular structure. 
In addition it is now clear that about 7 of these ring waves are influenced by interactions between low degree f-modes and g-modes, providing probes of static stability and composition gradients in Saturn's interior, a first in planetary seismology. 
The mechanism and utility of ring seismology for studies of Saturn's interior are reviewed in \cite{Mankovich20b}.

\begin{figure*}[ht]
	\centering
	\includegraphics[width=0.95\textwidth]{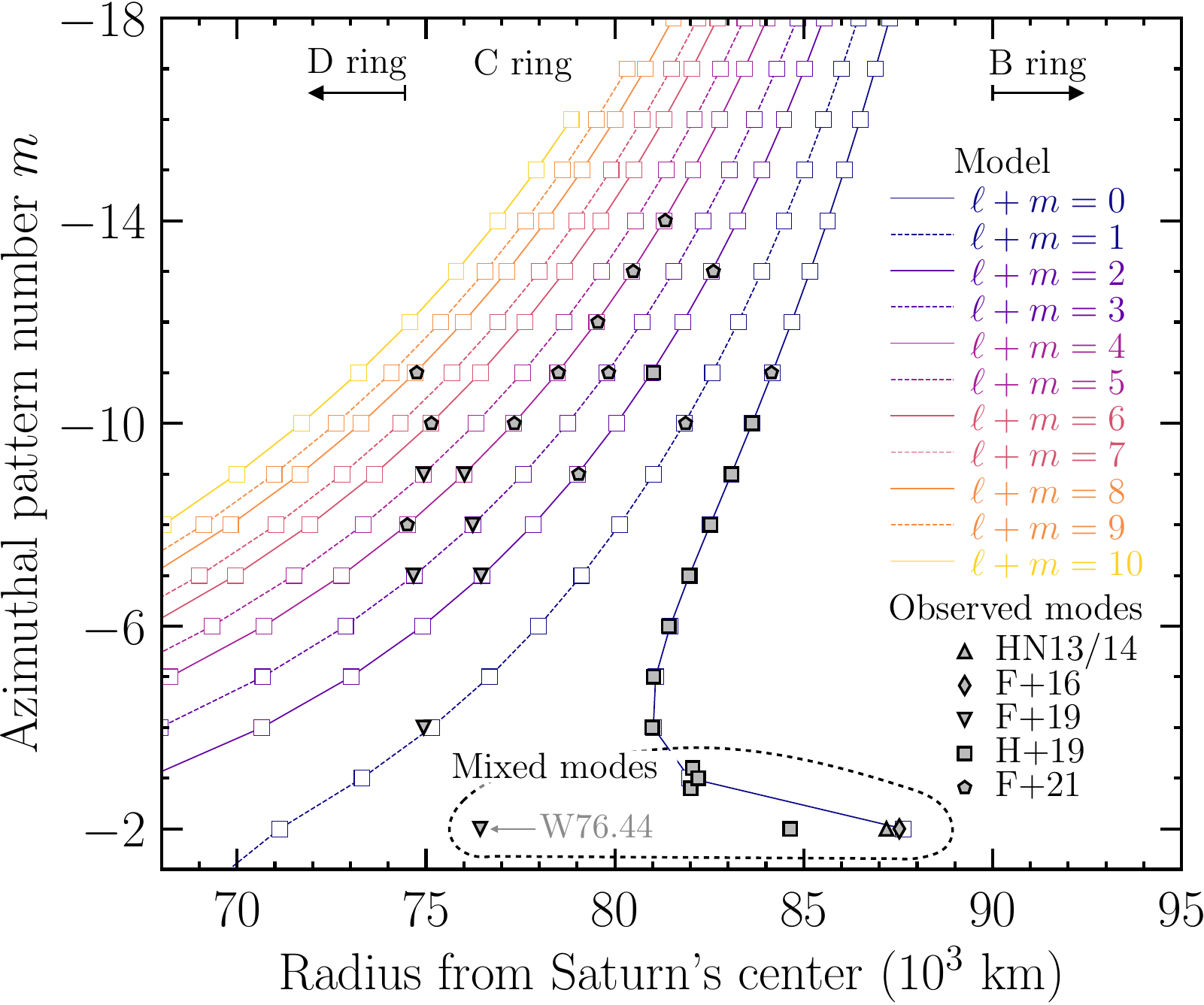}
    \caption{Observed wave patterns in Saturn's C ring (filled symbols) compared with expected resonance locations for a theoretical spectrum of Saturn f-mode oscillations (open boxes connected by colored lines). Modeled frequencies are shown for a three-layer interior model from \cite{Mankovich19} and are colored by the corresponding normal mode's latitudinal wavenumber $\ell+m$. Resonances falling in the tenuous D ring close to Saturn or in the optically thick B ring 
    are unlikely to be observed.
    \label{fig.ring_seis_summary}
    }
\end{figure*}

Each wave's location (or pattern speed) and $m$ value together precisely measure the oscillation mode frequency in the planet, which in in turn is sensitive to Saturn's rotation through a combination of the Doppler shift and intrinsic rotation effects (e.g., the Coriolis force).
Because the wavefunctions of distinct planetary oscillation modes are concentrated in different regions of the planetary interior, the full set of frequencies provides an exceptional new window into the variation of density, buoyancy, and rotation within Saturn's interior. 
Fig.~\ref{fig.ring_seis_summary} summarizes the known seismology constraints from the C ring, compared to the model of \citet{Mankovich19} who used the pattern frequencies of 14 waves to measure the solid body rotation rate of Saturn as $P_{\rm Sat}=\rm 10h\,33m\,38s^{+1m\, 52s}_{-1m\,19s}$. 

The spate of subsequent ring wave detections by \cite{French21} show remarkable agreement with predictions from this 2019 model (Fig.~\ref{fig.ring_seis_summary}) and extend the catalogue of Saturn's seismic constraints to angular degrees as high as $\ell=19$ and latitudinal wavenumbers as high as $\ell+m=8$.
With the seismic constraints now spanning a swath of radial and angular structures inside Saturn, it should be possible to further refine the seismic measurement of Saturn's bulk rotation and quantify departures from uniform rotation.
The full interpretation of this remarkable and growing dataset is still in its early stages; recent findings from modeling efforts are summarized in \S\ref{sec.seis_models} below. 


\subsubsection{He abundance}
The determination of Saturn's atmospheric helium abundance has long been a pressing question, as it has important implications for the role of hydrogen-helium phase separation in the planet's deep interior \citep{SS77b,FH03,Mankovich20a}.  Since there has not yet been a Saturn entry probe to measure atmospheric abundances in situ, helium abundance determinations have been done via remote sensing, which is a difficult data analysis problem.  A review of the \emph{Voyager}-era remote sensing issues, which typically combine radio occultation data and thermal infrared spectroscopy, is provided in \citet{Conrath2000}.

More recent determinations of Saturn's atmospheric He abundance have come from two papers.  The first \citep{Achterberg20} uses thermal infrared spectroscopy from \cas's Composite Infrared Spectrometer (CIRS) of wavelength regions that are sensitive to He-H$_2$ collision-induced absorption (CIA) features.  They determine a He/H$_2$ number ratio of 0.04--0.075, well below the protosolar mixing ratio, suggesting significant amounts of helium have rained down within the planet.  Alternatively, \citet{Koskinen18} use the \cas\ Ultraviolet Imaging Spectrograph (UVIS) (above the homopause) and limb scans observed by CIRS (below the homopause) to create empirical atmospheric structure models for Saturn (assuming an eddy diffusion coefficient constrained by the vertical methane distribution), to assess the atmosphere's mean molecular weight with height.  These authors find a volume mixing ratio of $0.125 \pm 0.025$, suggesting only modest He depletion.   The size of the error bar on both of these modern measurements, and more importantly, the large difference between them, leads to significant uncertainties in models of the planet's current structure and evolution, and stands in stark contrast with the precise determination of the He/H$_2$ in Jupiter's atmosphere of $0.157 \pm 0.003$. \citep{vonzahn98}.

\subsection{Interior structure models constrained by planet's gravity field} \label{gravitymod}

\subsubsection{Interior Model Background}

\begin{figure}[ht]
	\centering
	\includegraphics[width=0.48\textwidth]{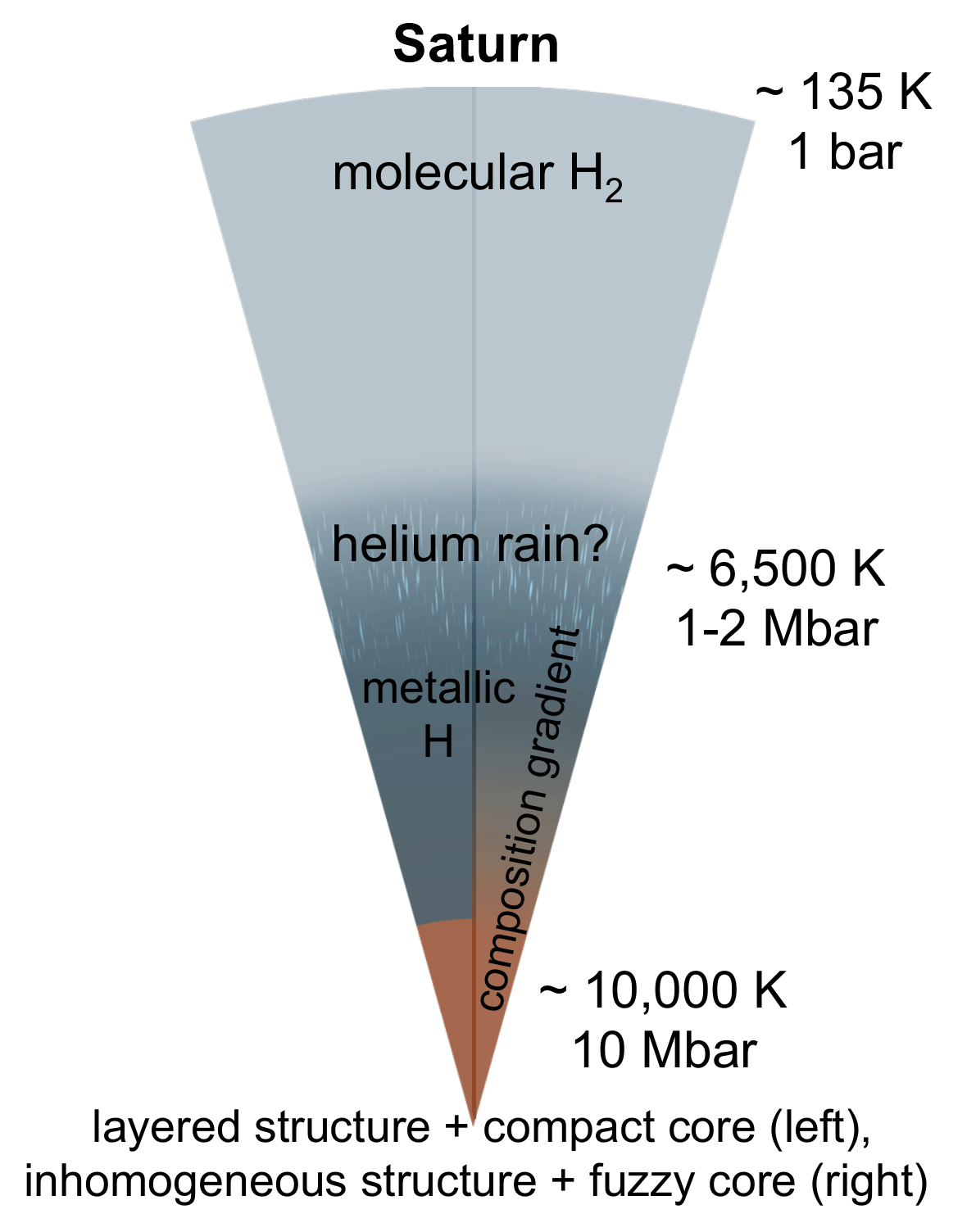}
    \caption{Sketches of Saturn's possible internal structure. Left: traditional layered structure. Right: Deep interiors with fuzzy core and composition gradients. In this case Saturn's deep interior is expected to be non-adiabatic (adapted from \citet{Helled2020}). 
\label{sketch}
}
\end{figure}

Historically, the interior structure of Saturn was divided into distinct layers as shown in Figure \ref{sketch} (left). At the center is a compact core composed predominately of rock and ice. The hydrogen-helium envelope is divided into an outer region, where hydrogen forms molecules, and an inner region where hydrogen is predicted to be a fluid atomic metal. The inner and outer envelope is separated by a layer in which helium is immisicible and rains out of the mixture  \citep{SS77b}. In some models \citep{Puestow2016,Mankovich20a}, the helium rain layer spans the entire inner envelope resulting in a helium ``ocean'' forming outside the core.

The heavy element core could exhibit some degree of mixing, referred to as a ``dilute'' or ``fuzzy'' core and characterized by a composition gradient extending outwards into the inner envelope (Figure \ref{sketch} right). {\it Ab-initio} materials calculations have demonstrated that both rock and ice phases are soluble in hydrogen at the conditions of Saturn's core \citep[][and references therein]{Gonzalez2014}. However, gravitational stability of the heavy core material could maintain a compact core against convective redistribution with the overlying envelope \citep{Moll17}.

In an attempt to be consistent with current (and still-evolving) usage, we here characterize cores as either ``compact'' or ``dilute/fuzzy.'' Compact cores consist of rocks, ice, or some combination thereof; their Z value (the heavy elements mass fraction) is 1.  The dilute/fuzzy core will have H/He ``mixed in," with $Z<1$, and the fuzziness could be a portion of the core, or the entire core, and this region may have a gradient in Z. Although there is not necessarily a well-defined boundary between a dilute/fuzzy core and the envelope, in some models its heavy element abundance is represented by a constant-Z value (Ni 2020), while in others by a Gaussian-Z distribution \citep{Wahl2017a}, or another kind of gradient \citep{Mankovich21}.  The latter, discontinuous case is equivalent to what has been denoted inner envelope metallicity \citep{Nettelmann13b} or deeper envelope \citep{miguel2016}.  For consistency with previous usage, in those cases (see Table \ref{tab:cores}, later) we tabulate as ``core mass'' only the mass of the compact core.  Otherwise, we count as core mass the mass of heavy elements in the dilute/fuzzy region plus the mass of a central compact core, if any.


Interior models of Saturn, with distinct layers similar to Figure~\ref{sketch} (left), have previously been fit to pre-Grand Finale \cas\ data \citep{HG13,Nettelmann13b}.  Updated modeling results using the \cas~Grand Finale gravity were presented in \citet{Iess2019} and explored in greater detail in \citet{Militzer19} and \citet{Ni2020}. For further modeling details on typical model construction, see \citet{Fortney2016}. 


Given our focus on \cas\ and space constraints of the chapter, we can only touch on equation of state (EOS) issues.  Recent tabulations of H/He EOS are found in \citet{Militzer2016b,Chabrier2019} and a recent review of hydrogen is found in \citet{Helled2020}.  To be clear, uncertainties in the equation of state directly translate into uncertainty of the interior structure models.  As an example of uncertainties still present in theory, we note that predictions from density functional theory (DFT) \citep[e.g.,][]{Militzer2013b}, more widely used in planetary EOS calculations, and Monte Carlo calculations by~\citet{2018PhRvL.120b5701M}, yield 5\% differences in the density of pure hydrogen at 6000~K over a wide pressure range.

Due to progress with laboratory experiments, the entire pressure range in Saturn's and Jupiter's interiors can now be probed with dynamic compression techniques~\citep{Kritcher2020}.  While it has remained a challenge to determine the density with an accuracy of 1\%, the real difficulty lies in the temperature measurements. In many shock wave experiments, temperature is not measured but inferred from models.

\begin{figure}[ht]
	\centering
	\includegraphics[width=0.5\textwidth]{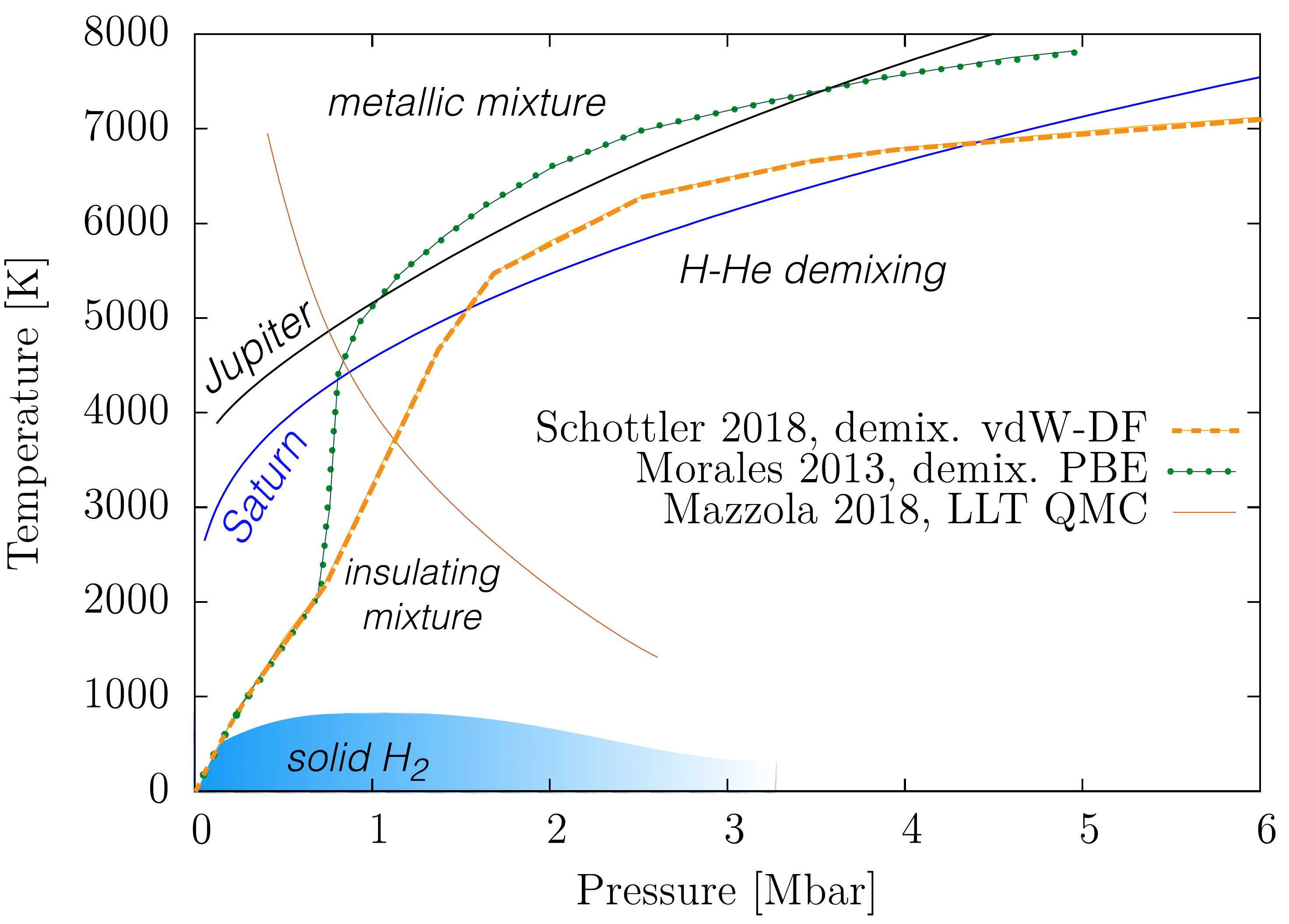}
    \caption{Phase diagram for a H-He mixture of a proto-solar composition as predicted by numerical calculations together with representative \emph{P-T} profiles of  Jupiter and Saturn. Green: Calculations from PBE with non-ideal entropy of mixing \citep{Morales09}. Orange: demixing region using the vdW-DF and non-ideal entropy of mixing \citep{Schottler18}. The red line corresponding to the metallization of H as predicted by \citet{2018PhRvL.120b5701M}.  Jupiter and Saturn isentropes are labeled. 
    The figure is modified from \citet{Helled2020}. 
\label{fig.imm}}
\end{figure}

Furthermore, the location of the hydrogen-helium immiscibility (``demixing") line in Fig.~\ref{fig.imm} is an additional source of uncertainty (see section~\ref{sec:HHe} and recent experimental work by \citet{Brygoo21}). Some uncertainties are introduced when different equations of state at low and high pressures are combined into one model but most of them can be avoided if absolute entropies~\citep{Militzer13} are used that enable one to determine the temperature for given entropy at different pressure points independently.  


To conclude, one has to keep in mind that all the interior models constructed using physical EOSs are affected by their uncertainties. As a result, the inferred composition depends to some degree on the choice of the EOS. Therefore, in order to advance our understanding of the giant planets progress in experiment and computer simulations of relevant materials at high pressure and temperature is required. 


Since planets cool by convection, Saturn's envelope is most often assumed to have an adiabatic temperature profile. Under this assumption, the temperature through the outer envelope is defined by an isentrope of the chosen equation of state with entropy $S_{\rm mol}$ chosen to be consistent with the observed temperature at 1 bar.  Typically this is cited at $135 \pm 5$K.  In practice, though, \citet{Militzer19} use 143 K after correcting the 135 K value from~\citet{Lindal1992} for a change in (very uncertain) helium abundance~\citep{G99}.  However, regions with static, stable compositional gradients, or exhibiting double-diffusive convection (sometimes referred to as semi-convection) can have super-adiabatic gradients ~\citep{Leconte13,Nettelmann15}.

The helium rain layer is one example of a region of possible non-adiabatic behavior, while the core region is another. Different implementations of helium rain in Saturn based on various phase diagrams have been adopted \citep{FH03,Puestow2016,Militzer19,Mankovich20a}.  Within a framework of a model that matches the gravity field, any He abundance less the protosolar value in the outer regions must be made up for with an enrichment at depth, which conserves total helium.

\subsubsection{Matching interior models to gravity measurements} 

The plausibility of a given interior structure model is judged by translating the composition and temperature parameters for each layer into a density distribution, from which the gravity field can be calculated and compared to spacecraft measurements. It is not possible to uniquely infer the composition and the distribution of the elements. As a result, the exact composition and internal structure of Saturn remains unknown. However, more accurate gravity data can be used to further constrain the interior. 
Prior to the \cas~Grand Finale, the best measurements of Saturn's gravity \citep{Jacobson2006} constrained only the first three even zonal harmonics $J_2$, $J_4$ and $J_6$.
As summarized in Section \ref{sec.grav_obs} the gravity observations \citep{Iess2019} from \cas's Grand Finale orbits improved constraints on $J_2$-$J_6$ and determined higher order moments of Saturn's field, thus placing much stricter constraints on the planet's interior
density distribution.

The unprecedented precision of \cas~ Grand Finale gravity measurements, as well as \juno's measurements at Jupiter, necessitated the development of more accurate methods for calculating self-consistently the shape and gravity field \citep{Hubbard2013, Wahl2017b, Militzer19, Nettelmann21}. 

A planet's density profile influences the shape of the gravity field through rotation and tides, which cause equipotential surfaces within the planet to deviate from spherical symmetry. In the case of the rapidly rotating Jovian planets, the non-spherical potential is dominated by the centrifugal force. For a planet with a uniform rotation rate (i.e. no internal dynamics from winds) both the density distribution and resulting gravity field
are axisymmetric and north-south symmetric, meaning that only even zonal harmonics, $J_{2n}$, are non-zero. It is important to note, however, that even for such a simplified case the relationship between $J_n$ and a particular density distribution is inherently non-unique \citep{Movshovitz22}.

For a uniformly rotating planet in hydrostatic equilibrium, the
magnitudes of the even zonal harmonics decay as $|J_{2n}|\sim q_{\rm rot}^n$, where
$q_{\rm rot}$ is the ratio of the centrifugal and gravity accelerations at the
equator. As illustrated in Fig.~\ref{fig.JupiterSaturnJn} the gravity field of Jupiter measured by the \juno~spacecraft \citep{Folkner2017a} is broadly consistent with this relationship. By contrast, the observed Saturn field exhibits significant divergence from the expected relationship for even moments $J_6$ and higher (see Tab.~\ref{tab:Jn}). \citet{Iess2019} and \citet{Galanti19} demonstrated that
Saturn's observed gravity cannot be reproduced with models that assume uniform rotation, and
instead requires deep differential rotation \citep{Hubbard1982}.

\begin{figure}[ht]
	\centering
	\includegraphics[width=0.50\textwidth]{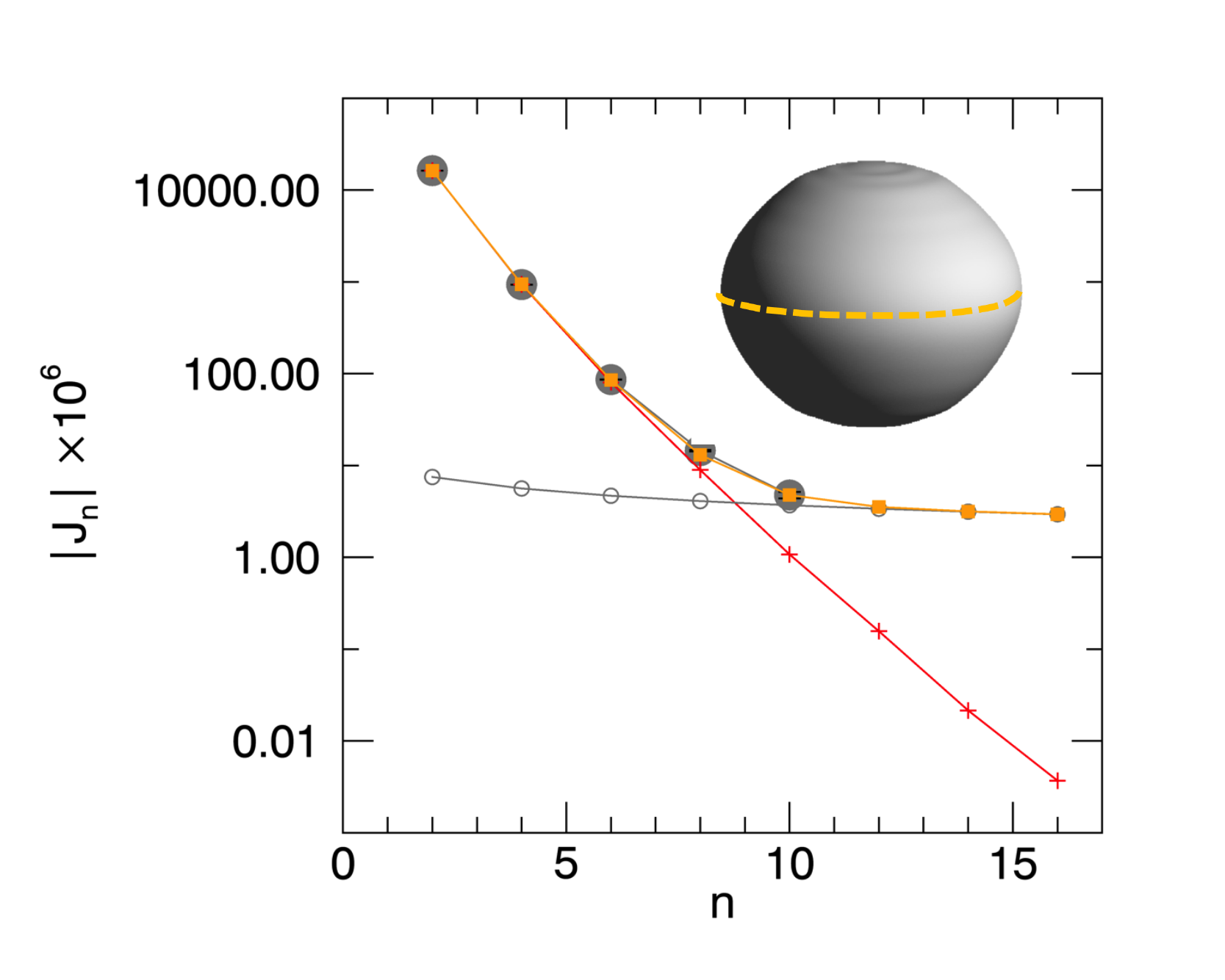}
    \caption{Thin-cord model for Saturn's high-order gravity harmonics. The red pluses are $J_n$ calculated for a uniform rotating planet, the grey open circles are the $J_n$ from thin cord, while orange squares are the combined gravity signal.  The grey filled circles are the \emph{Cassini} measurements (plotted up to $J_{10}$) all larger than the (vertical) error bars, for clarity.}
      \label{fig.ThisCordModel}
\end{figure}

This departure can be illustrated in terms of the simple thin-cord model shown in
  Fig.~\ref{fig.ThisCordModel}.  Here we have a Saturn model of mass $M_S$ in hydrostatic equilibrium rotating with a single
  rotation period $P_{\rm Sat}$ (shaded inset),   The gravitational moments of this Saturn
  model are shown as red pluses, falling to unobservable small values at $J_{12}$ and
  beyond.  We imagine this Saturn model's equator to be encircled by a very thin
  cord of total mass $\mu \ll M_S$, shown as a dashed line on the shaded inset.
  Using Eq. (3) and Eq. (6) of \cite{1978SvA....22..362T} 
  , we have for the contribution of this cord to the gravitational moment $J_n$,   
   $$
   \Delta J_n=-\frac{\mu}{M_S} P_n(0),
   $$
   where $P_n(0)$ is the value of the n-th degree Legendre polynomial evaluated at argument 0, the equator. The
   $\Delta J_n$ for the thin cord, assuming $\mu / M_S = 1.5 \times 10^{-5}$, are plotted
   as open circles in Fig. \ref{fig.ThisCordModel}.  The total observed $J_n$ are then the sum of the solid-body
   $J_n$ (pluses) and the $\Delta J_n$, as shown in Fig. \ref{fig.ThisCordModel}, qualitatively reproducing the
   observed values.
   
   The postulated thin cord has a gravitational signature similar to
   that of an equatorial uplift induced by
   strong eastward zonal flow near the equator; we note that the best-fit value for
   $\mu / M_S $ is orders of magnitude larger than estimated values for Saturn's
   total ring mass, so the rings cannot possibly account for the progression in $J_n$.

\subsubsection{Oblateness and rotation period} \label{sec.interior.rotation}


Images of the giant planets Jupiter and Saturn show a banded texture, which is associated with winds of latitude-dependent velocities. In contrast, their deep interiors are commonly thought to rotate rigidly. 
This rigid body rotation rate is an important parameter for models of the giant planets as Jupiter and Saturn in several aspects. It defines the reference system for the observed winds and determines the 
centrifugal forces acting on the rotating planet. The stronger such force, the more oblate the planetary shape becomes. As shape and gravity field are intermingled, any inference of internal structure through gravity field observations, such as acquired during the Cassini Grand Finale, ultimately depends on the uncertainty in rotation rate.
Nevertheless, Saturn's degree of flattening (oblateness) offers an independent constraint on the rotation rate when combined with interior structure models \citep{Helled09}.

In contrast to Jupiter, where magnetic field observations from space allowed to infer Jupiter's rotation period to high accuracy of less than 0.01 s, the increasingly 
accurate Saturn magnetic field observations culminating in the Cassini Grand Finale revealed an increasingly close spin axis---magnetic dipole axis alignment, preventing determination of its rotation rate by magnetic field observations. Nevertheless, an interplay of multiple Saturn (system) observations and model predictions has, at about the time of the Cassini Grand Finale, been converging toward a value around 10h 33m. 

Several observables related to Saturn's magnetosphere have periods, ranging from the Voyager value of 10h 39m 22s inferred from radio emissions to 10h 47m 6s inferred from non-axisymmetric components 
of Cassini magnetic field data \citep{Giampieri06}. However, due to the time-variability of the magnetic field data over rather short time-scales of days to decades, as revealed by Cassini, none of these periods 
necessarily reflects the rotation of the deep interior. Alternative methods predict faster rotation.

One method to predict a significantly faster rotation of 10h 32m is the minimization method, where the dynamical height of the 100 mbar surface is minimized by either using zonal wind or oblateness observations \citep{Anderson07}. This method reproduces the system III rotation period of Jupiter to within 1 minute \citep{Helled09}. 
Another method relies on shape and gravity data. Using Pioneer and Voyager occultation data and Cassini Grand Finale-like tight constraints on the low-order gravity field, the rotation period was predicted to fall within 10h 31m and 10h 38m \citep{Helled2015}. 

By combining radio occultation oblateness measurements \citep{Lindal1985}, Cassini grand finale gravity data on $J_2$--$J_{10}$, and interior models, the rotation rate was constrained to within 1 minute around 10h 33m 34s \citep{Militzer19}. 

\begin{figure}[ht]
	\centering
	\includegraphics[width=0.48\textwidth]{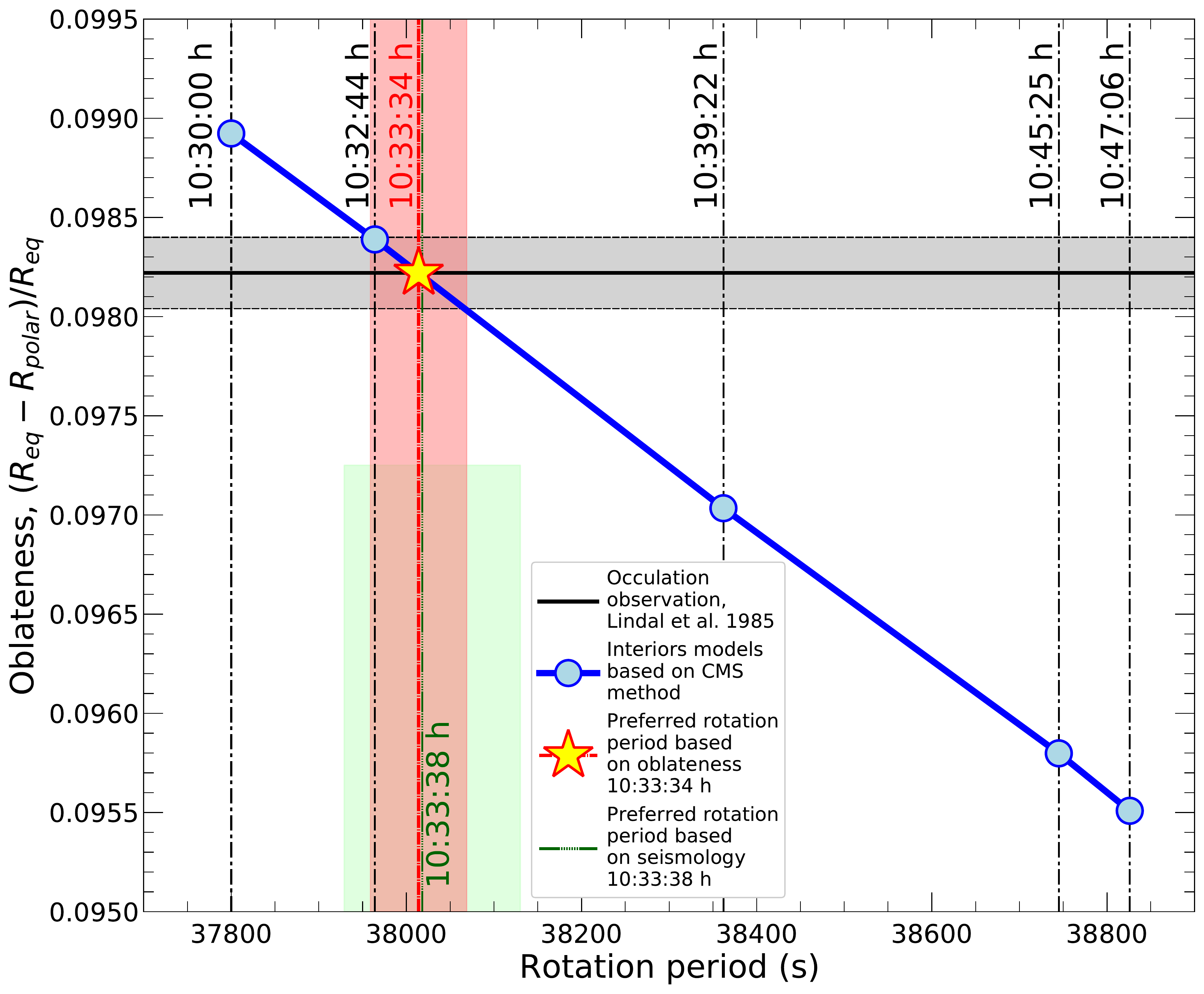}
    \caption{Oblateness derived from CMS models with different rotation
      periods compared to the radio occultation oblateness measurement by
      \citet{Lindal1985}, which determined an oblateness of 0.09822 $\pm$
    0.00018. Based on this comparison, the derived a rotation period of
    10:33:34 h $\pm$ 55 s \citep{Militzer19} (red with shaded error) this is consistent with the rate of 10:32:45  h $\pm$ 46 s, inferred by~\citet{Helled2015} and the one estimated from ring seismology of 10:33:38$\,$h$^{+112\rm s}_{~-89 \rm s}$ \citep{Mankovich19} (green with shaded error).
\label{fig.period}}
\end{figure}

Figure ~\ref{fig.period} compares Saturn interior models inferred by \citet{Militzer19} with differential rotation constructed for five different rotation periods ranging from
10:30:00 to 10:47:06 h. For each period, it is possible to construct
interior models, including differential rotation on cylinders, that match all even $J_{n}$. However, only a narrow range in rotation period
is consistent with the \citet{Lindal1985} radio occultation measurements. A similar analysis performed on the identical radio occultation measurements with solid-body rotation \citep{Anderson07}, suggests that the contribution of differential rotation leads to a $\sim$50 s slower prediction using this method.

The \citet{Militzer19} estimate is in excellent agreement with the results of \citet{Helled2015} who used gravity and shape data and calculated a period of  10:32:45 h $\pm$ 46 s, and with the results based on the analysis of ring seismology, with a period of 10:33:38$\,$h$^{+112\rm s}_{~-89 \rm s}$ \citep{Mankovich19}, which is further discussed in Section \ref{sec.seis_models.rotation}.





\subsubsection{The influence of rotation period on other properties}
\label{sec.interior.rotation_influence}

The assumed rotation rate influences the inferred heavy element mass fraction in the H/He envelope $Z$, with faster rotation allowing for higher atmospheric $Z$-values but also a stronger possible heavy element gradients $Z(r)$ toward the center \citep{Nettelmann13b}. Using \cas~Grand Finale gravity data, up to $4\times$ solar atmospheric $Z$ is found \cite{Militzer19}. While this value is higher than that of comparable models for Jupiter,  which struggle to pass beyond $1\times$ solar, still falls short of the observed methane enrichment of 9$\times$ solar, as was already discussed based on pre-Finale data \cite{Fortney2016}.

\begin{figure}[ht]
\includegraphics[width=0.48\textwidth]{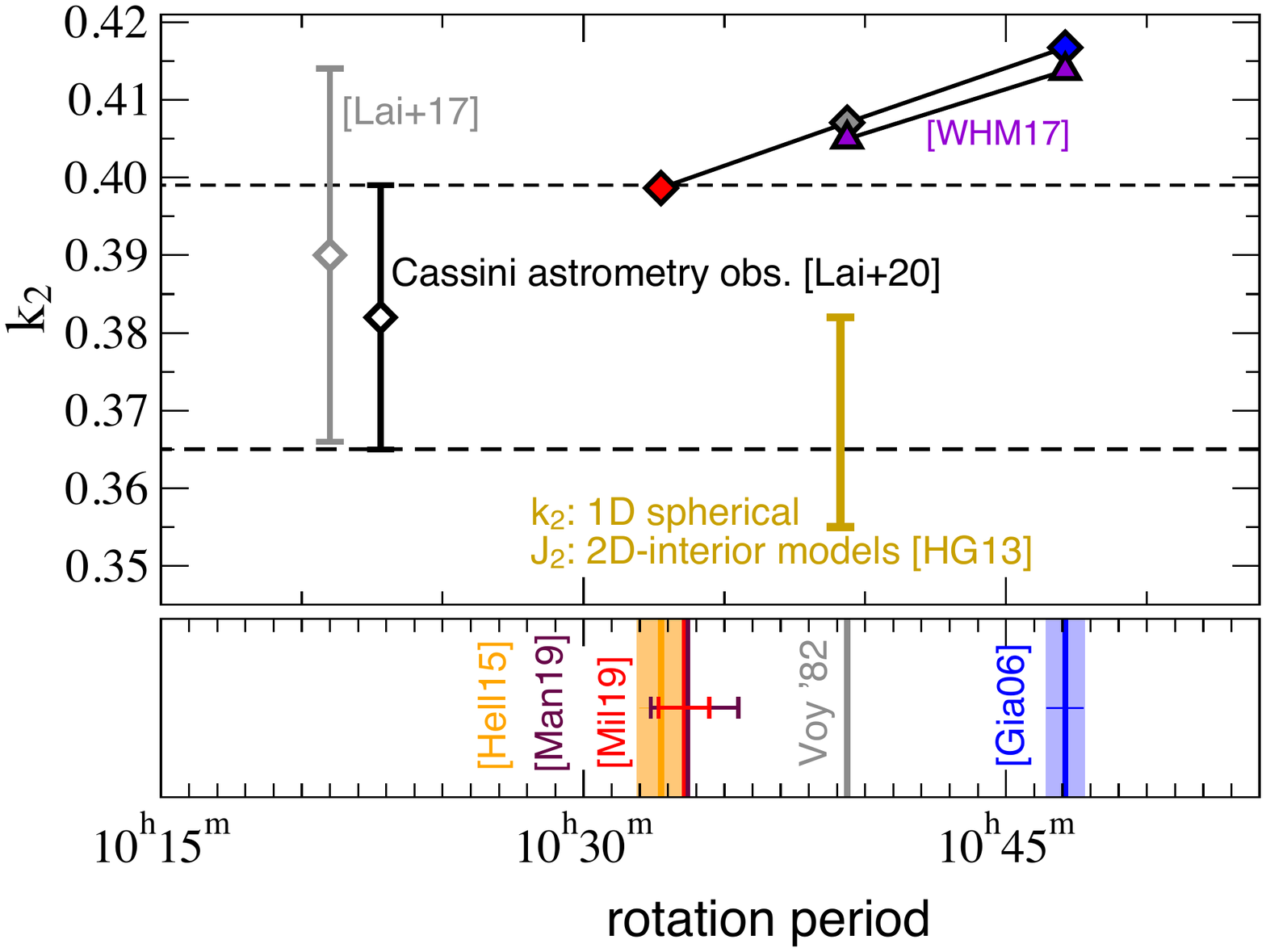}
\caption{Love number $k_2$ values and Saturn's rotation period. Top: observationally derived (open symbols) $k_2$ values of \citet{Lainey2017} and \citet{Lainey20} with $3\sigma$ formal uncertainty (dashed horizontal lines) placed arbitrarily on x-axis;  calculated static $k_2$ values of 3D Saturn models (colored triangles: \citet{Wahl2017b}, colored diamonds: 2D interior models of \citet{Nettelmann13b} fed into the 3D modeling tool of \citet{Nettelmann19}), vertical yellow bar: $k_2$ of 2D models in \citet{Lainey20}. Model $k_2$ values depend on assumed rotation period. Bottom: predicted rotation rates from magnetic field (grey, blue), winds and shape (orange), oblateness (red), and seismology-related ring features (dark red).} \label{fig:k2_Prot}
\end{figure}

Beyond interior models and wind models, the rotation rate is an important parameter; it also influences the moment of inertia and the Love numbers $k_{nm}$, which quantify the planet's tidal response to its satellites or parent star. A lesson learned from Juno at Jupiter is that superposition of dynamic effects changes its $k_{22}$ value by $\sim 5\%$ compared to the static value \citep{Durante20}. While Jupiter's \emph{observed} $k_{22}$ value is inferred from spacecraft tracking (Juno), for Saturn it is observations of its moons' positions acquired over decades.  With \cas, that allowed for a first estimate of Saturn's $k_{22}$ value, yielding a value of $0.390\pm 0.024$ \cite{Lainey2017}. Further spacecraft tracking of \cas\ and astrometric/imaging data allowed \cite{Lainey20} to refine ${\rm Re}(k_2)=0.382$ with a $3\sigma$ formal uncertainty of $\pm0.017$ (Figure \ref{fig:k2_Prot}.)

Faster rotation favors lower static $k_{22}$ values from interior model predictions. Notably, a rotation rate of 10h 34m or faster, as suggested by the diverse independent approaches, brings the static model $k_{22}$ value into agreement with the measured one, but  the central value is not reached by any static tidal response model \citep{Wahl2017b} unless the rotation period would be further lowered to $\sim$ 10h 15m. However, given the uncertainty in Saturn's $k_{22}$ value, which is due to uncertainties in the tidal state of Enceladus but also due to differences in numerical orbit predictions, it might be too early to suggest that dynamic effects play a similarly small but non-negligible role on Saturn as on Jupiter. Figure \ref{fig:k2_Prot} illustrates the available Saturn rotation periods and $k_2$-estimates.

\subsubsection{Inferred Depth of Differential Rotation} \label{sec.diff_rot}

Large-scale zonal wind patterns in Saturn's upper atmosphere have been characterized and monitored using optical cloud tracking both 
preceding and for the duration of the \cas~mission \citep{GM2011}. The most prominent feature of the observed wind profiles is a pronounced eastward
jet centered on the equator. Prior to the \cas~Grand Finale gravity
measurements, the depth of these zonal flows was unconstrained, and their effect was not considered in
earlier modeling studies of Saturn's interior~\citep{HG13,Nettelmann13b}. 

Although the possible contribution of deep winds to Saturn's gravity is a highly non-unique problem in principle, with the Grand Finale gravity measurements it is possible to test a model in which the cloud-level velocity profiles are mapped onto a hypothesized deep flow pattern.
When the zonal-wind velocity profile is extended to depths of many scale heights, it modifies the axisymmetric
gravitational field, meaning that the even $J_{2n}$ are modified from the values expected for
a uniformly rotating body in a predictable fashion \citep{Hubbard1982,Hubbard99b}. Since the velocity profiles of the observed winds have a north-south asymmetry, there must also be a small asymmetric contribution to the gravity, manifesting itself as non-zero
odd $J_n$ \citep{Kaspi2013}. \citet{Iess2019} reported measurements $J_3$ and $J_5$ that are thus interpreted as a consequence of differential rotation.

\citet{Iess2019} explores two different methods for incorporating differential rotation into gravity models.
The first treats interior dynamics purely as uniform rotation on cylinders that extend all the way through the planets equatorial plane.
This simplified flow structure has the benefit that it can be described using potential theory \citep{Hubbard1982}, and thus can be integrated into the the CMS gravity simulation in a fully self consistent manner \citep{wisdom2016}. The contribution to the potential from differential rotation modifies the shape of the equipotential
surfaces, which feeds back into the calculated gravitational field. The downside of this method is that it cannot account for a decay of velocity with depth, which is expected due to interactions with the magnetic field as hydrogen's
electrical conductivity increases with increasing pressure~\citep{Cao2017}.

\begin{figure}[ht]
	\centering
	\includegraphics[width=0.48\textwidth]{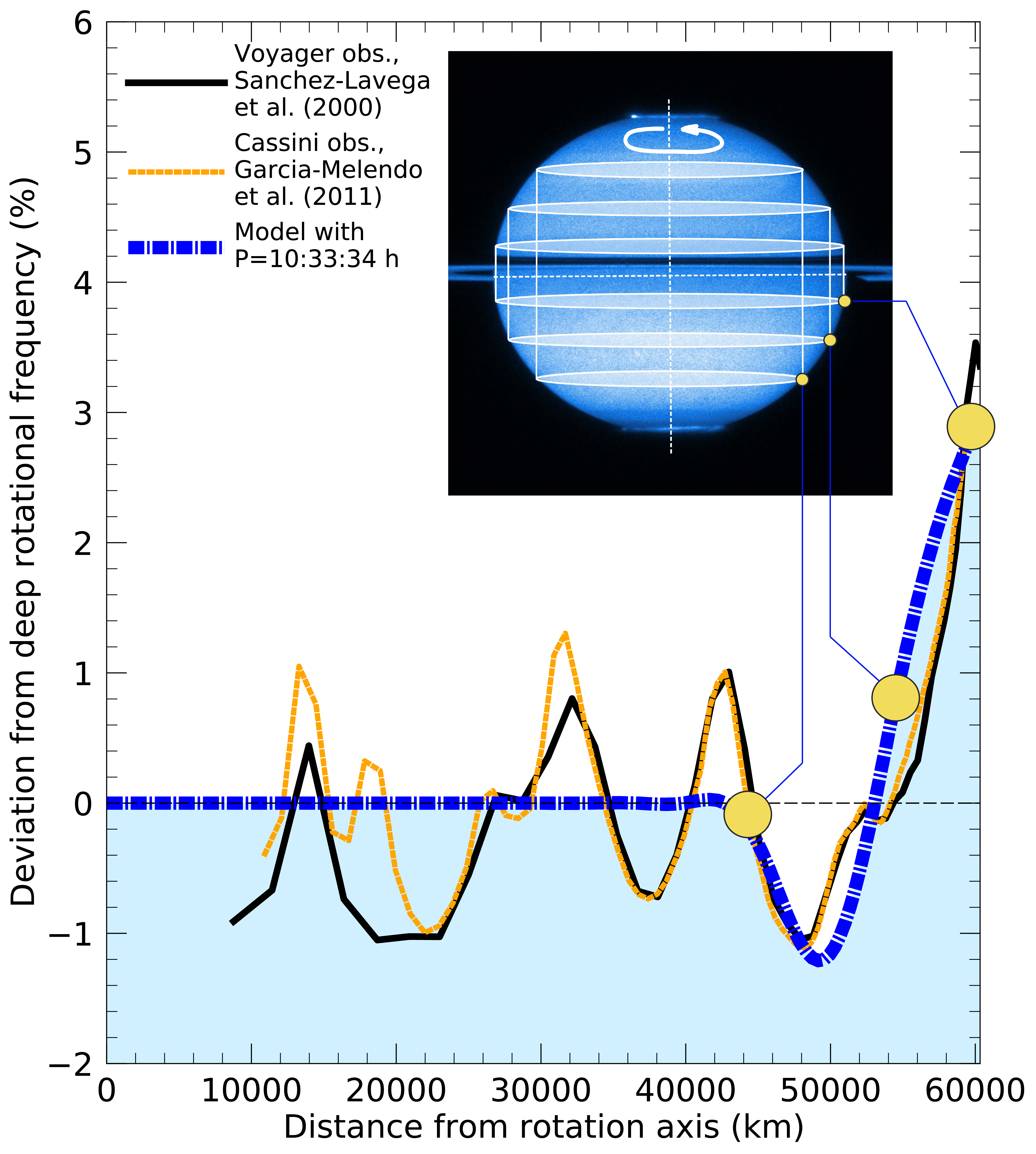}
    \caption{Average of the rotation profiles in a suite of Saturn
  interior models that match the observed even gravity harmonics. Differential rotation must be several thousands of
  kilometers deep. Rotation on cylinders reproduces the eastward equatorial jet
  that rotates about 3\% faster than the deep interior. The inset shows an
  illustration of the cylinders. The rotation frequencies inferred by
  tracking the clouds in Saturn's visible
  atmosphere~\citep{GM2011} are shown for comparison.
\label{fig.diff_rot}}
\end{figure}

Figure \ref{fig.diff_rot} shows a velocity profile for differential rotation on cylinders, optimized to match the even $J_{2n}$ from the \cas~Grand Finale, compared with time and latitude averaged wind velocities from cloud tracking \citep{GM2011}. The good agreement between the  rotation on cylinders model and the observed winds beyond $\sim$45,000 km from the rotation axis suggest that the zonal winds extend to depths exceeding 10,000 km, and that the dynamic contribution to the gravity is dominated by the equatorial jet. 
Rotation on cylinders cannot adequately represent the high-latitude winds, since at high latitudes the winds must be truncated, rather than extending all the way through the center of the planet. Differential rotation on cylinders is also north-south-symmetric by definition, so the odd $J_n$ cannot be reproduced by this method.

Figure \ref{fig.diff_rot} is a corrected version of Fig.~5 of \citet{Militzer19}, where the
observed cloud-tracking wind speeds were incorrectly referenced to a frame rotating
with a period of 10:33:34~h rather than the System III period of 10:39:22.4~h.  When this
adjustment is made, lowering the ordinates of the cloud-tracking data by about one percent,
we obtain Fig.~\ref{fig.diff_rot}, with improved agreement with the cloud-tracking data. As before the model reproduces the eastward equatorial jet that rotates faster than the deep interior. But now, the models also reproduce the westward jet that rotates 1\% slower than the deep interior. The CMS models in \citet{Militzer19} cannot reproduce the jets at higher latitudes because differential rotation on cylinders was assumed and they would need to reach unphysically deep into the planet's interior to reproduce the high-latitude jets. If one instead terminates these cylinders at finite depth, the jets at all latitudes can be reproduced reasonably well \citep{Iess2019} as is demonstrated with the following approach.

The second method assumes the winds introduce a correction to the gravity solution obtained from a uniformly rotating model. The thermal wind
equation~\citep{Kaspi2013,Galanti2017,Galanti19} or the gravitational thermal wind
equation~\citep{kong2013} are employed to derive a local correction to density from the assumed flow structure. This density change is integrated to find corrections to the gravitational moments. This method allows for consideration of more complicated wind fields at the expense of introducing additional approximations, and neglecting the feedback of differential rotation on shape of the planet's internal density distribution. \citet{Iess2019} includes calculations in which the observed  $J_n$, including the odd moments, are
matched by optimizing a function describing decay of the surface wind velocities with depth. For these models, wind depths of $\sim$9000 km best match the \cas~Grand Finale gravity, in rough agreement with the prediction from rotation on cylinders.

For comparison, the depth of the winds in Jupiter was recently determined thanks to the Juno mission \citep{Iess2018,2018Natur.555..223K,2018Natur.555..227G}. The depth was estimated to be 2000--3000 km, suggesting that 1\% of the outer part of Jupiter (by mass) is  differentially rotating in patterns similar to the observed atmospheric winds. 
This penetration depth is in agreement with the one inferred using theoretical calculations of the ohmic dissipation constraint associated with the metallization of hydrogen \cite[e.g.,][]{2008Icar..196..653L}, but see also \citet{Christensen20} for possible complications. 

It is expected that the winds can penetrate deeper regions in the case of Saturn since hydrogen metallization occurs deeper due to its smaller mass and resulting pressures \citep{2017Icar..296...59C}.
For Jupiter and Saturn the depths of the winds  correspond to about 95\% and 80\% of the total planetary radius,  respectively.  
A detailed discussion on that topic can be found in \citet{2020SSRv..216...84K}. 

\subsubsection{Results from recent Structure Models} \label{sec.predict}

\subsubsubsection{Evaluation of interior parameters} \label{sec.predict2}

\begin{figure*}[p]
	\centering
	\includegraphics[width=0.75\textwidth]{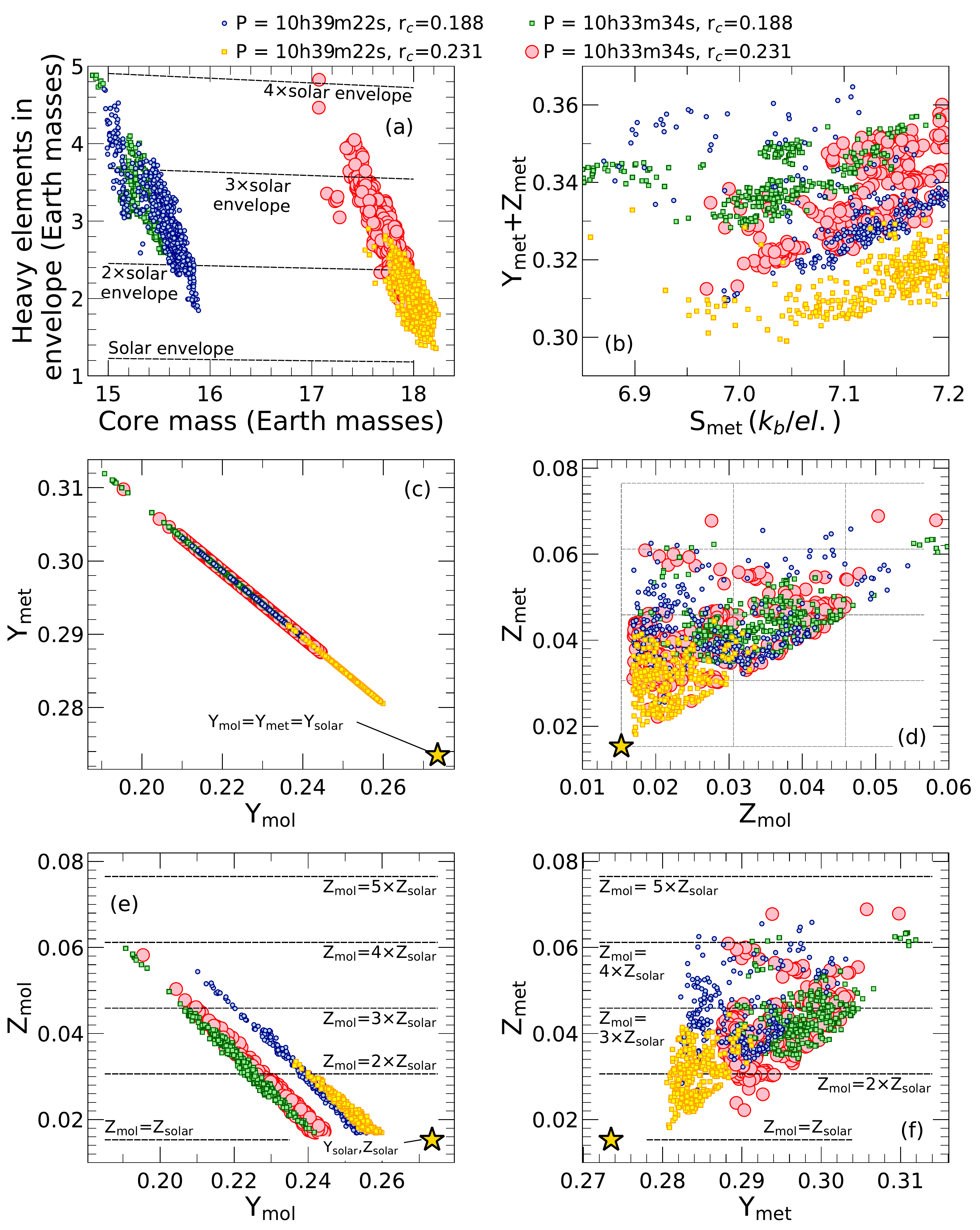}
  \caption{
    Comparison of parameters from interior models with the preferred rotation
      rate of 10:33:34 h from \citet{Militzer19}, compared to the System III
      rotation rate.
      Core radii $r_{c}=0.188$ and $r_{c}=0.231$ correspond to rocky and rock-ice compositions respectively.
      (a) The distribution of heavy element mass between the core and envelope, suggesting around 20 \me\ of heavy elements in total.
      (b) The variation of the mass fraction elements heavier than hydrogen with
      entropy in the inner, metallic envelope.  Higher internal entropies leader to warmer, less dense hydrogen, which can accommodate more heavy elements.
      (c) The variation of helium mass fraction in the molecular and metallic envelopes.  The Sun's bulk value is 0.275.
      (d) The variation of heavy element mass fraction in the molecular and metallic envelopes.  The value in the Sun is about 0.015.
      (e) The tradeoff between heavy element and helium mass fractions in the
      molecular envelope.  As $Y$ decreases, $Z$ is able to increase.
      (f) The tradeoff between heavy element and helium mass fractions in the
      metallic envelope.
      In panels (b), (c), (d), (e), and (f) solar values are shown with a
      yellow star, corresponding to an assumed end-member case with no partitioning
      of helium through rain-out.
      \label{fig.para}
    }
\end{figure*}

Figure~\ref{fig.para}, adapted from \citet{Militzer19}, shows the interior model parameters from ensembles of
models assuming distinct layers, generated with Monte Carlo sampling for two different rotation periods and core radii. 
Here a core radius of $r_c$=0.188 corresponds to a compact core with an iron-silicate (0.325:0.675) composition, while
$r_c$=0.231 corresponds to a solar iron-silicate-water ice (0.1625:0.3375:0.5) composition, assuming homogeneous mixtures 
 using the additive volume rule with their respective equations of state. Panel (a) shows the distribution of heavy element mass between the envelope and core.  
\citet{Militzer19} infers the mass of Saturn's
core to be $\sim$15--18 \me\ with additional heavy
elements (1.5--5 \me) distributed throughout the planet's H/He envelope. This is in good agreement with 4-layer Saturn models of \citet{Ni2020} who obtains 12--18 {\me} for the accumulated mass of heavy elements in a compact core and an overlying diluted core layer.

Panels (b-f) of Figure~\ref{fig.para} show the relationship between the compositional parameters ($Y$ and $Z$) in the outer (mol) and inner (met) envelope, along with the entropy $S$ as a proxy for envelope temperature, for the same ensembles of interior models. Models with longer rotation periods are found to be more restricted in parameter space, requiring higher interior entropies, lower envelope metallicities and/or lesser degree of helium rainout. The models with shorter rotation periods produce $Y_{\rm mol}$ that are compatible with reanalyzed Voyager measurements of atmospheric helium, $\sim$0.6--0.8$\times$solar \citep{Conrath2000}, while models with longer rotation rates require less depletion of helium in the outer envelope. The
\cas~Grand Finale gravity permits models without enrichment of heavy elements in the inner envelope ($Z_{\rm mol}\approx Z_{\rm met}$). By contrast interior models of Jupiter fit to \juno~gravity data require either an enrichment of $Z$ across the helium rain layer \citep{miguel2016} or an extended `dilute' core \citep{Wahl2017a}, neither of which is required for Saturn by gravity alone.

The core masses of $\sim$15--18 \me\ found from \citet{Militzer19} models that use \cas~Grand Finale gravity fall within the range of core masses predicted from earlier modeling efforts, which ranged from 0-25 \me \citep{Fortney2016}.  The significantly improved gravity field certainly better constrains the interior, however, the derived core mass in a structure model also depends sensitively on the modeling choices or philosophy.  For instance, previous models considered a wider range of inhomogeneous $Z$ distributions, \citep{HG13} or explored different equations of state \citep{Nettelmann2012}, compared to \citet{Militzer19}.

Unknowns of the deep interior likely make the uncertainty in the core mass larger than a few \me.  For instance, if a larger fraction of Saturn's interior is stably stratified as indicated by the interpretation of ring oscillations (Section 4), perhaps a relic of planet formation (Section 6), the deep interior could be warmer and thus be more enriched that inferred in models that only account for super-adiabaticity in the helium rain region \citep{Militzer19,Ni2020}, where super-adiabaticity may be low \citep{Mankovich20a}.  
The \citet{Militzer19} models all assume a compact core,  an assumption that is in contrast with results from formation models. As a result, the actual core size and the heavy element mass in Saturn's deep interior could be different if dilution of the core material by the hydrogen helium envelope is considered \citep[e.g.,][]{Wahl2017a,Mankovich21}.  Lastly, several Earth masses of He could have rained down to the core, adding to the best-fit heavy element core mass \citep{Mankovich20a}.

In the H/He envelope, The 1--4$\times$solar uniform enrichment of all
heavy elements in the models of \citet{Militzer19} is lower than would be expected from  observations atmospheric methane concentrations, consistent with $\sim$9$\times$ solar enrichment of
carbon \citep{Fletcher2009}.  For Saturn, atmospheric temperature and composition have never been measured {\it in situ}. If the envelope is hotter than expected, it could trade off
for higher concentrations of heavy elements, without significantly
affecting other model predictions. Measurements of other
heavy elements include both much lower (N/H
$\sim$3$\times$ solar) and higher S/H $\sim$13$\times$ solar) concentrations, although these differences might reflect model dependence in determining abundances, or from measuring
regions of the atmosphere that are not well mixed~\citep{Atreya2019}. 

For Jupiter, most elements measured by the Galileo Entry Probe showed an enrichment in comparison to solar by about a factor of three (see chapter by Atreya et al.). This may imply that Jupiter's outer envelope is enriched in comparison to a solar composition.  On the other hand, structure models of Jupiter that are based on state-of-the-art EOSs find solutions where Jupiter's atmospheric metallicity is about solar \citep{Wahl2017a}, significantly lower than the Galileo probe.  For the water, current estimates based on microwave radiometer (MWR) data of the equatorial region suggest that oxygen is 1-2 solar \citep{Li20}. This information would further constrain structure models and in addition would provide critical information linked to the origin and evolution of the planet \cite[e.g.,][]{2014MNRAS.441.2273H,Vazan2018}. 

For both Jupiter and Saturn, if observed atmospheric element abundances are taken as representative of the entire H/He envelope, or even lower bound for the envelope $Z$, there appears to be a ``heavy element crisis'' in both Jupiter and Saturn, as structure models yield lower values in both planets.  The implication of this situation is still being actively debated and models with more complex metal enrichments as a function of depth are being explored \citep{Debras_2019}. As a result, linking the measured atmospheric with internal structure models is challenging and it is unclear to what degree the they can be used to further constrain the interior. 


\subsubsubsection{Constraints from models with deep subsurface differential rotation} 

The Saturn models of \citet{Militzer19} described above assume that the observed zonal flows can be extended downward to a certain depth. \citet{Militzer19} varied internal heavy element abundances, which mainly influence $J_2$ and $J_4$, and the wind speeds of the cylindrical flows until good agreement with the observed flows and all even moments was achieved. This approach led to the finding of $J_2$ and $J_4$ being barely affected by the winds, while the winds would modify $J_6$ by a large amount (compared to Jupiter) of 6\% \citep{Iess2019}. However, these Saturn models --like the vast majority of planet models for gaseous planets-- assume a spheroidal shape derived under the assumption of an equipotential surface although the observed winds are inconsistent with that assumption.

\citet{Ni2020} adopted a broader approach by also allowing for winds that differ at depth from the surface winds. Such winds are predicted if the assumption of spheroidal shape is relaxed \citep{Kong2019}. Moreover, subsurface winds are found to extend to greater depth, up to 20,000 km \citep{Kong2019}, and thus have a non-negligible influence also on $J_2$ and $J_4$ \citep{Ni2020}. Nevertheless, the resulting heavy element abundances from these models are generally in good agreement with those of \citet{Militzer19}. Notably, Ni (2020), as previously found by \citet{Saumon04} and \citet{Nettelmann13b}, finds envelopes that are homogeneous in $Z$ over a wide range of atmospheric He abundances and rotation rates. Despite this wide parameter range, the obtained atmospheric metallicty falls within 0--8$\times$ solar, which is lower than the observed C/H ratio in Saturn's atmosphere, with a core mass estimate of 12--18 {\me}.

\subsubsubsection{Constraints from investigations that minimize modeling assumptions} 

\begin{figure}[ht]
	\centering
	\includegraphics[width=0.50\textwidth]{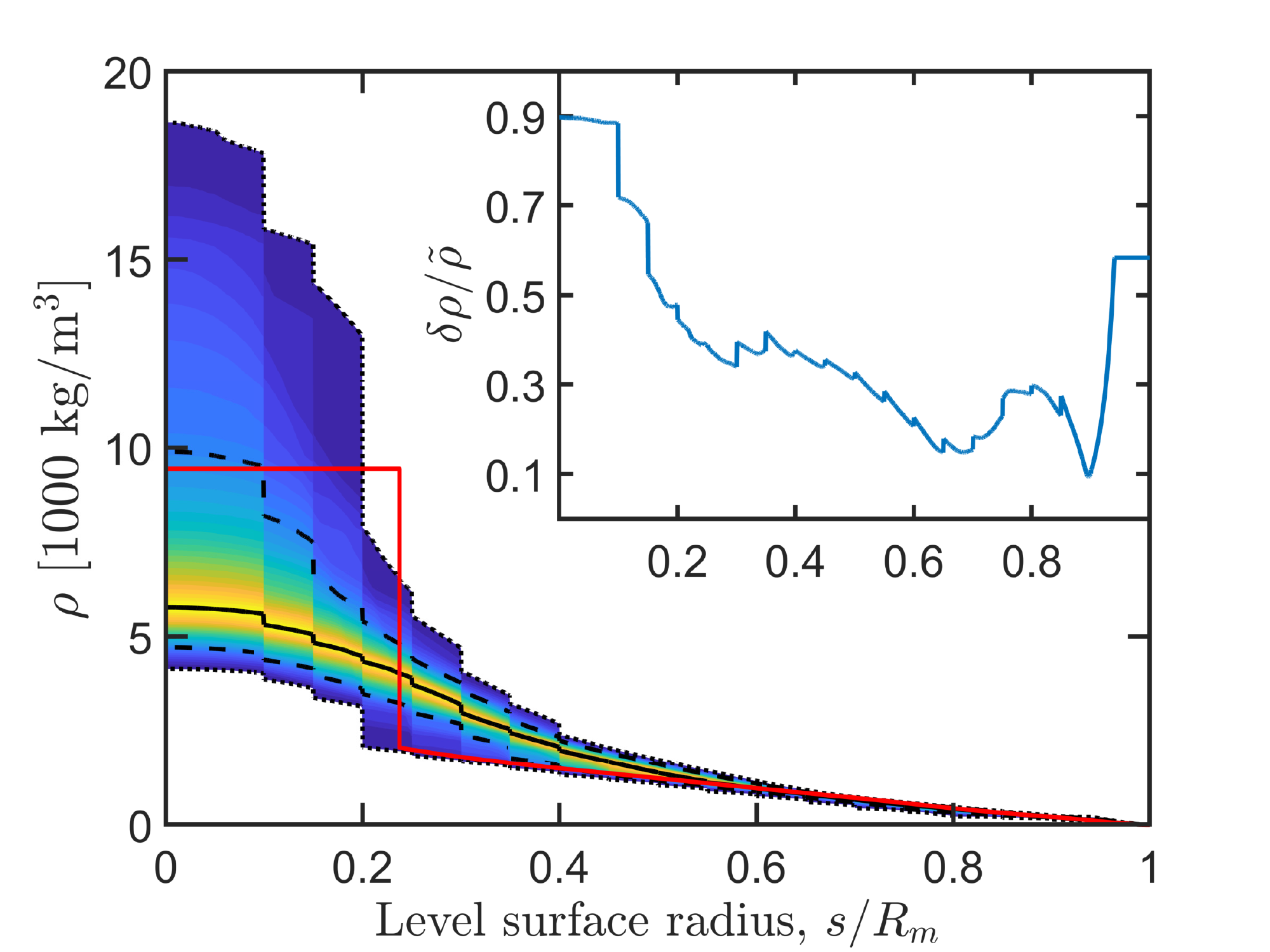}
    \caption{The posterior probability distribution of Saturn interior density profiles as inferred from empirical models. The density profile is represented by polynomials. The thick black line corresponds to the the sample-median of density on each level surface. The dashed lines mark the the 16th and 84th percentiles and the dotted lines mark the 2nd and 98th percentiles; between the lines percentile value is indicated by color.  A model from \cite{Militzer19} is shown in red, which also fits the gravity data, but is outside of the most likely region from \citet{Movshovitz2020}. The inset shows the relatively uncertainty on the density profile as a function of radius, which is less well constrained near the surface (where there is little mass) and in the deep interior (where gravity data does not probe well). 
\label{figemp}}
\end{figure}

Standard internal structure models of Saturn (and gaseous planets in general) depend on model assumptions, in particular the assumed composition and the EoS used by the modeler. 
An alternative and complementary approach to model planetary interiors is to represent the density profile by a mathematical function or other approaches. Then one searches for a density profile that reproduces all the observations without assuming a specific composition and number of layers \cite[e.g.,][]{Helled2009,Movshovitz2020}.  A recent study by \citep{Movshovitz2020} used polynomials to represent Saturn's interior. They used  Markov-Chain Monte Carlo in order to map all the possible interior density profiles of Saturn. 
Figure \ref{figemp} shows the density profiles that match Saturn's gravity field and other physical properties. While it is clear that it is not possible to uniquely determine Saturn's density profile, some key conclusions can be made.  
It was found that the outer half of Saturn's radius is relatively well constrained, as the gravity field probes the outer half the planet's mass distribution much better than the inner half.  The density structure of the outer layers was interpreted as a significant metal enrichment which is consistent with atmospheric measurements. 

The deep interior of Saturn is harder to constrain, but the models indicate the existence of a ``core.'' Note, however, that this inferred core has rather low densities ($\sim$ 6 g cm$^{-3}$) suggesting that Saturn's deep interior consists of a fuzzy core and/or composition gradients, although the statistical significance of this finding is not high.  Figure \ref{figemp} serves as a visual reminder that core properties are not well determined by gravity-field data alone.
 
\subsection{Models constrained by ring seismology data} \label{sec.seis_models}

\subsubsection{Evidence for stable stratification}
\label{sec.seis_models.stable}
Normal mode seismology is unique in its ability to probe interior composition gradients directly.
In a convective environment, the Brunt-V\"ais\"al\"a frequency $N$
characterizing the buoyant response of the fluid is imaginary, corresponding to convective instability: small, adiabatic, vertical fluid displacements grow exponentially.
The case of $N$ identically zero corresponds to the limit of perfectly adiabatic convection, an idealization that describes Saturn's interior extremely well unless composition gradients or opacity windows \citep{Guillot95} are present.
If in an otherwise convective environment one introduces a stabilizing (bottom-heavy) gradient in heavy elements or helium, then $N$ takes real values corresponding to the \emph{oscillatory} response of the fluid to adiabatic vertical displacements.
These oscillatory motions are internal gravity waves, which manifest rather dramatically in the case of ring seismology at Saturn: some initially puzzling constraints from \cas\ data \citep{Hedman13}---visible at $m=-2,-3$ in Fig.~\ref{fig.ring_seis_summary}---are now understood as the product of the prograde Saturn f-modes anticipated by \cite{Marley90,Marley91} and \cite{Marley93} coupling to a spectrum of g-modes hosted in the deeper interior \citep{Fuller14}.
Fuller proposed that a stable stratification between Saturn's core and envelope gives rise to g-modes whose frequencies overlap with low-degree $(\ell=2,3)$ f-modes.
The result is a series of mixed modes that could explain the unexpectedly dense spectrum of patterns observed in the rings at low azimuthal orders ($m=-2,-3$). \cite{French16} bolstered this hypothesis by detecting an additional $m=-2$ density wave in the Maxwell ringlet, confirming a prediction from \cite{Fuller14}.

While a major conceptual step forward in light of the early \cas\ ring wave detections, Fuller's interior model was ultimately a simplistic polytrope-based model only roughly constrained by Saturn's $J_2$.
These basic ideas would face the test of new seismic constraints from subsequent wave detections, new gravity constraints from the \cas\ Grand Finale, and more systematic study using interior models based on physical equations of state.

Among the multitude of more recent wave detections associated with Saturn oscillations (\citealt{French19}, \citealt{Hedman19}, \citealt{French21}), only a single $m=-2$ wave called W76.44 in the inner C ring appears to be associated with g-modes and thus Saturn's composition stratification.
\cite{Mankovich21} created new models based on the \cite{Militzer13} ab initio hydrogen EoS, featuring a single stably stratified region connecting an rock/ice-rich inner core to a hydrogen/helium-dominated envelope, and found W76.44 to be associated with the lowest radial order  ($n=1$) $\ell=2$, $m=-2$ g-mode of Saturn.
This mode resembles a quasi-interface mode trapped on the stratified core-envelope transition.
Consequently its frequency was observed to be remarkably sensitive to the width of Saturn's stably stratified core-envelope transition, a fact that could not be exploited by \cite{Fuller14} when this pattern had not yet been discovered.

\begin{figure}[ht]
	\centering
	\includegraphics[width=\columnwidth]{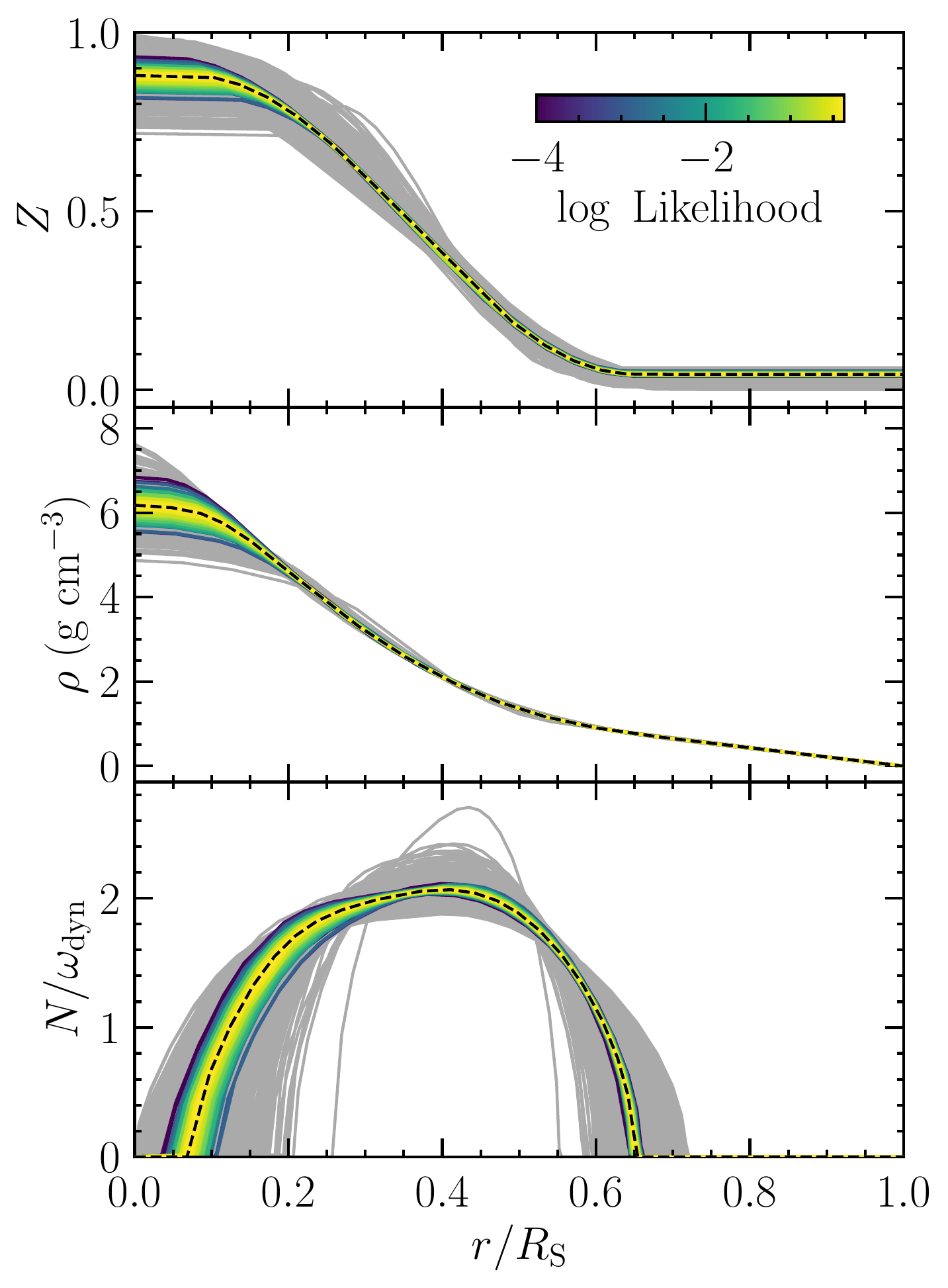}
    \caption{Interior models inferred from ring seismology and \cas\ Grand Finale gravity. 
    Profiles colored by their likelihood are for a three-parameter interior model; the best of these is superimposed in dashed black curves. 
    The broader distribution of grey profiles are for a seven-parameter interior model.
    The top and middle panels show the heavy element mass fraction and mass density as a function of mean radius; the bottom panel shows the Brunt-V\"ais\"al\"a frequency normalized by Saturn's dynamical frequency $\omega_{\rm dyn}\equiv(GM_{\rm S}/R_{\rm S}^3)^{1/2}$.
    Figure taken from \cite{Mankovich21}.
    \label{fig.ring_seis_models}
    }
\end{figure}

Figure~\ref{fig.ring_seis_models} displays the main results of the \cite{Mankovich21} models, obtained from joint fits to the $m=-2$ ring seismology and the zonal harmonics $J_2$, $J_4$ and $J_6$ from the \cas\ Grand Finale.
With central densities of roughly $6\ {\rm g\ cm}^{-3}$ these models are very similar to the most likely empirical models obtained by \cite{Movshovitz2020} (see Fig.~\ref{figemp}) but are considerably more tightly constrained, a combination of the added seismic constraints and the stronger set of assumptions involved in EoS-based models with particular parameterizations of the composition structure.  

The main finding of this joint retrieval is that in all cases the stably stratified region is large, encompassing about 60\% of Saturn's interior by radius (Figure~\ref{fig.ring_seis_models}).
This dilute core model is consistent with the composition gradient case sketched in Fig.~\ref{sketch}.
Such a broad transition proved necessary in order to produce the modest typical stratification $N/\omega_{\rm dyn}\sim2$ implied by the g-mode frequencies observed in the rings, while still maintaining the significant core-envelope density contrast required by the even zonal gravity harmonics.

An implication of this model that warrants some skepticism is that the highly electrically conductive inner regions are found to be stable to overturning convection, at odds with the simplest view that Saturn produces its magnetic field through a ``standard'' dynamo in a fully convective metallic hydrogen interior.
In reality, a stable stratification over the inner 60\% of the planet's radius does not spell doom for its magnetic field generation. First, 
a sizable fraction of the volume exterior to $\sim0.6\,R_{\rm S}$ is strongly electrically conductive (see Cao et al.) and so field generation by a dynamo process in a shell outside the stratified region is possible.
Second, the seismology only requires that the inner 60\% by radius be stably stratified in a radius-averaged sense: the constraining modes are g-modes with radial wavelengths on the order of $R_{\rm S}$, and hence are not greatly sensitive to fine spatial structure. This means that the observations could also be consistent with a more complicated composition profile, such as a convective staircase \citep{Wood13} extending to $0.6\ R_{\rm S}$, provided that the convective layers remain small compared to the g-mode radial wavelength \citep{Belyaev2015}.
It is thus possible for an atypical dynamo process to take place in convective layers embedded within a metallic hydrogen interior that is nonetheless stably stratified on average.
As discussed in Section \ref{magmag} and the chapter by Cao et al., the magnetic power spectrum observed by \cas\ betrays a complicated structure that may require multiple dynamo surfaces or multiple stable regions, and so we can say with certainty that the structure presented in Fig.~\ref{fig.ring_seis_models} is an oversimplification.

\subsubsection{Constraints on rotation}
\label{sec.seis_models.rotation}

The remainder of the new C ring wave detections are patterns with higher azimuthal orders $|m|\geq4$, where the f-modes are trapped closer to the planetary surface and hence naturally decouple from the deep g-modes \citep{Fuller14,Mankovich21}. 
As a result these new constraints say little about the deep interior but are useful probes of Saturn's rotation, which is itself a major contributor to uncertainty in the interior structure (see Sections~\ref{sec.interior.rotation} and \ref{sec.interior.rotation_influence}).
Fig.~\ref{fig.ring_seis_summary} compares the full set of ring waves associated with Saturn oscillations to the prograde f-mode frequencies for a simplistic interior model (three distinct layers, each homogeneous in its composition) constrained by $J_2$ and $J_4$.
\citet{Mankovich19} applied a family of such models and found that the overall fit to the f-mode frequencies hinged strongly on the rotation rate assumed for Saturn's interior.
Rotation enters the observable ring pattern frequencies via the Doppler shift from the Saturn frame, and to a lesser extent through intrinsic changes to mode frequencies due to the Coriolis force and higher order rotation effects terms modifying the response of the fluid.
Putting aside the $m=-2,-3$ waves complicated by g-modes, Mankovich et al. used the remaining 14 f-mode frequencies that were known at the time to estimate Saturn's bulk rotation period as $P_{\rm Sat}=\rm 10h\,33m\,38s^{+1m\, 52s}_{-1m\,19s}$.
While not a tighter constraint than all previous and contemporary estimates \citep[e.g.,][]{Read09,Helled2015,Militzer19}, it is remarkable that this independent and rather direct probe of Saturn's rotation, possible only in the era of \cas, delivered a consistent result.

Fig.~\ref{fig.ring_seis_summary} shows that the most recent ring detections \citep{French21} agree extremely well with the \cite{Mankovich19} predictions. 
However, this broad-scale comparison conceals small but significant residuals: these ring resonances are located to a relative precision of order $10^{-5}$ in location or pattern speed, while the modeled pattern frequencies exhibit relative departures from these at a level of $10^{-4}$ to $10^{-3}$.
The mismatch furthermore appears to depend systematically on the angular degree $\ell$ of the f-mode in question \citep{French21}, indicating that the underlying model is not fully capturing Saturn's true density or rotation profile.
Indeed all of the models discussed so far treat only rigid rotation, a picture now clearly ruled out by Grand Finale gravity.
The finer-scale interpretation of the ring seismology constraints for models with realistic flow profiles is an ongoing effort.

\subsubsection{Synthesizing gravity science and ring seismology}
\label{sec.seis_models.gravity}

No structure model that has been put forward for Saturn is satisfactory in all respects. 
Based on existing modeling results, the \cas~gravity alone does little to constrain the details of the deep interior, beyond requiring $\sim$15 \me\ of heavy elements be concentrated near the center of the planet. Thus, models with a deep stably stratified layer \citep{Mankovich21} are likely permitted, but are in no way required by the \cas~Grand Finale gravity alone. 
Conversely, works prioritizing good agreement with Grand Finale gravity \citep{Militzer19,Ni2020} so far assume layered structures that are not supported by ring seismology constraints. 


To summarize, the Saturn mode frequencies measured by Cassini ring science present three pillars of support for a large-scale composition gradient supporting g-modes:
\begin{enumerate}
    \item The fine ($\lesssim1\%$) frequency splitting near the $m=-2$ and $m=-3$ f-modes requires a dense spectrum of ``additional'' oscillation modes with frequencies similar to the f-mode frequencies $\omega_f/\omega_{\rm dyn}\sim\sqrt\ell$.
    \item No non-f-mode detections have been made for $\ell\geq4$ (see Fig.~\ref{fig.ring_seis_summary}), implying that the ``additional'' modes only extend up to frequencies $\omega/\omega_{\rm dyn}\sim\sqrt{4}=2$, a signature of g-modes in a model with $N_{\rm max}/\omega_{\rm dyn}\sim2$. 
    \item The additional $m=-2$ wave W76.44 is located some 10,000 km away from the other $m=-2$ waves, a separation in line with the expected $\ell=2$ g-mode period spacing for Saturn models with $N_{\rm max}/\omega_{\rm dyn}$ of order a few.
\end{enumerate}
Despite the zonal gravity harmonics not requiring a dilute core in Saturn, these seismic constraints do appear to require one. 

However, the published models oriented toward joint seismology/gravity fits have their own limitations, chief of which is their strong assumption of a single region over which $Z$ and $Y$ both vary with prescribed functional forms; as we have seen, the reality is certainly more complicated.
The \cite{Mankovich21} models are far from unique, and it will be essential to study more general interior structures that decouple the heavy element gradient from the helium gradient, and allow for more than one stable region.

Some room for improvement of these seismology-forward models is that they have so far not accounted for Saturn's deep winds in a self-consistent way. Although the winds make negligible contributions to the frequencies of the \emph{isolated} low-degree modes that drove the main results of \cite{Mankovich21} regarding deep structure, the winds can play a strong role in the fine low-degree frequency splittings and certaintly in the frequencies of the more superficially confined high-degree f-modes \cite{2021PSJ.....2..198D}. Finally the perturbative treatment of Saturn's rotation adopted in most of the seismology modeling efforts so far 
limits the accuracy of predicted mode frequencies. This approach is just beginning to be supplanted by non-perturbative techniques \citep{2021PSJ.....2..198D} that will be essential for Saturn seismology in the longer term.

As is clear, even with the impressive gains in gravity and seismology data, there is no single model that matches all available constraints.  Given the non-uniqueness of gravity data, and the still developing understanding of seismology data, one model is not an achievable goal. Constraints on the total heavy element mass within the planet, and the core mass, depend significantly on the modeling framework chosen, including whether the interior profile is close to adiabatic, whether the core is compact or dilute, and more generally how helium and heavy elements are distributed within the planet. For reference, in Table \ref{tab:cores} we show  modeling compilations of planetary parameters. Recent models show a clear preference for a central heavy element concentration of $\sim$12-24 \me, with relatively little heavy elements in the H/He envelope. The interested reader must, however, delve into the details of each of the papers to understand why these differences arise, which we have done our best to briefly summarize.

\begin{table*}[h]
    \begin{center}
\begin{tabular}{| c | c | c | c | c|} 
\hline
Model reference   & Total heavy  & Core  mass     & Core & Core\\
 & element mass  (M$_\oplus$)     &  (M$_\oplus$) & radius ($R_{\rm s}$) & definition\\
\hline
\citet{HG13}    &        ~~5 $\dots$ 35 & 5 $\dots$ 20  & $\sim$ 0.2 & {}\\
\citet{Nettelmann13b} & $17\ldots27$ & $0\ldots 20$ & 0--0.2 & compact \\ 
\citet{Leconte12} &      26 $\dots$ 50     &   10 $\dots$ 21& 0.2 $\dots$ 0.3 & {}\\
\citet{Vazan2016} (model $S_3$)    &       $\sim$ 30  & 12   & 0.23  & dilute ($Z>0.5$)\\ 
Fig.~\ref{fig.para} adopted from \citet{Militzer19} & 18 $\dots$ 22 & 15 $\dots$ 18 & 0.19 $\dots$ 0.23  & compact \\
\citet{Movshovitz2020} &  $24\ldots 32$ & -- &  -- & \\
\citet{Ni2020} & $16.5\ldots 19.5$ &  $12\ldots 18$  & -- & compact $+$ dilute\\
\citet{Mankovich21} & $23\ldots26$ & $20\ldots24$ & $0.33\ldots0.36$ & dilute ($Z>0.5$)\\
\hline
\end{tabular}
    \end{center}
    \caption{Inferred total heavy-element mass, core masses, and core radii from a variety of Saturn interior models.  See the referenced papers for necessary modeling details.}
    \label{tab:cores}
\end{table*}

\subsection{Clues from satellite migration}
Measurements of Saturn's tidal response have led to unique, if less readily interpreted, constraints on interior structure.
Using a combination of astrometry and precision tracking of Titan's orbit using \cas\ range-rate data, \cite{Lainey20} reported a rapid orbital expansion indicative of very strong tidal dissipation inside Saturn.
The radiometric data enabled a precise determination of the imaginary part of Saturn's Love number ${\rm Im}(k_2)=-(69\pm57)\times10^{-4}$ at Titan's tidal frequency.
The implied quality factor $Q\sim100$, some two orders of magnitude smaller than the value expected from equilibrium tidal theory, lends overwhelming support to the theory that Titan's orbital evolution is dictated by resonance locking with low-frequency oscillations inside Saturn \citep{Fuller16,Lainey2017}.
For Enceladus, Tethys, Dione, and Rhea only weaker astrometric constraints are available, but still their corresponding estimates of $Q$ and orbit evolution timescales similarly favor the resonance locking hypothesis. In this picture the moons' orbits expand on Saturn's evolution timescale, tracking the spectrum of Saturn oscillation frequencies that in turn change with Saturn's evolving structure.

These planetary oscillations could take the form of inertial waves intrinsic to any Saturn model \citep[including conventional layered models; see][]{Ogilvie2004,Wu2005b} or, if Saturn does host a stable stratification, mixed inertial/gravity waves \citep{Fuller16}.
The basic mechanism thus does not strictly require static stability $N>0$, but having one generally increases the likelihood of resonant capture, since in that case mixed inertial/gravity waves would offer up a denser and broader spectrum of resonances capable of ensnaring satellite orbits \citep{Fuller16,Andre2019}.

\subsection{Results from Thermal Evolution Models} 

Models of the thermal evolution of giant planets aim to match a few particular quantities, typically the planet's \teff\ (or \tint) and radius at the planet's known mass and age.  The planet's current energy balance is an important aspect.  The total thermal luminosity yields \teff, but since $T_{\mathrm{eff}}^4 = T_{\mathrm{eq}}^4 + T_{\mathrm{int}}^4$, and since \teq\ depends on the planet's measured Bond albedo, the \emph{intrinsic} component of thermal flux (rather than re-radiated solar flux) is the important quantity.

A thermal evolution model aims to understand the secular cooling of the planet either from some initial formation model, or, more commonly, for an arbitrary hot initial state, not tied to any particular formation mechanism.  The radiative atmosphere serves as the bottleneck for cooling of the interior \citep{Hubbard99,Fortney11}. Potential problems with arbitrary initial conditions are discussed in \citet{Marley07} in the context of formation energy lost in the formation process. Other processes linked to the planetary formation that can affect the thermal evolution include the physics of the accretion shock \citep[e.g.,][]{2018MNRAS.477.4817C} and  the existence of composition gradients  \citep[e.g.,][]{Vazan2016}.

\subsubsection{Models with H/He phase separation}
\label{sec:HHe}
\citet{Mankovich20a} coupled the thermal evolution of both Saturn and Jupiter to a variant of the \emph{ab initio} He/H phase diagram of \citet{Schottler18}.  Since Jupiter and Saturn must obey the same physics, the evolution of both planets were treated as a coupled problem for the first time.  The phase diagram of \citet{Schottler18} was allowed to ``float" in temperature-pressure space, in terms of onset temperature for phase separation, and models needed to match \teff\ and radius for each planet and Jupiter's accurately measured $Y_{atmos}$.  Given the complexity of the phase space a Monte Carlo approach was employed, marginalizing over temperature shifts of the phase diagram, Saturn's uncertain Bond albedo, and any effects due to superadiabatic temperature gradients in regions with a He composition gradient \citep{SS77b,FH03,Nettelmann15,Mankovich16}.

\begin{figure}[ht]
	\centering
	\includegraphics[width=\columnwidth]{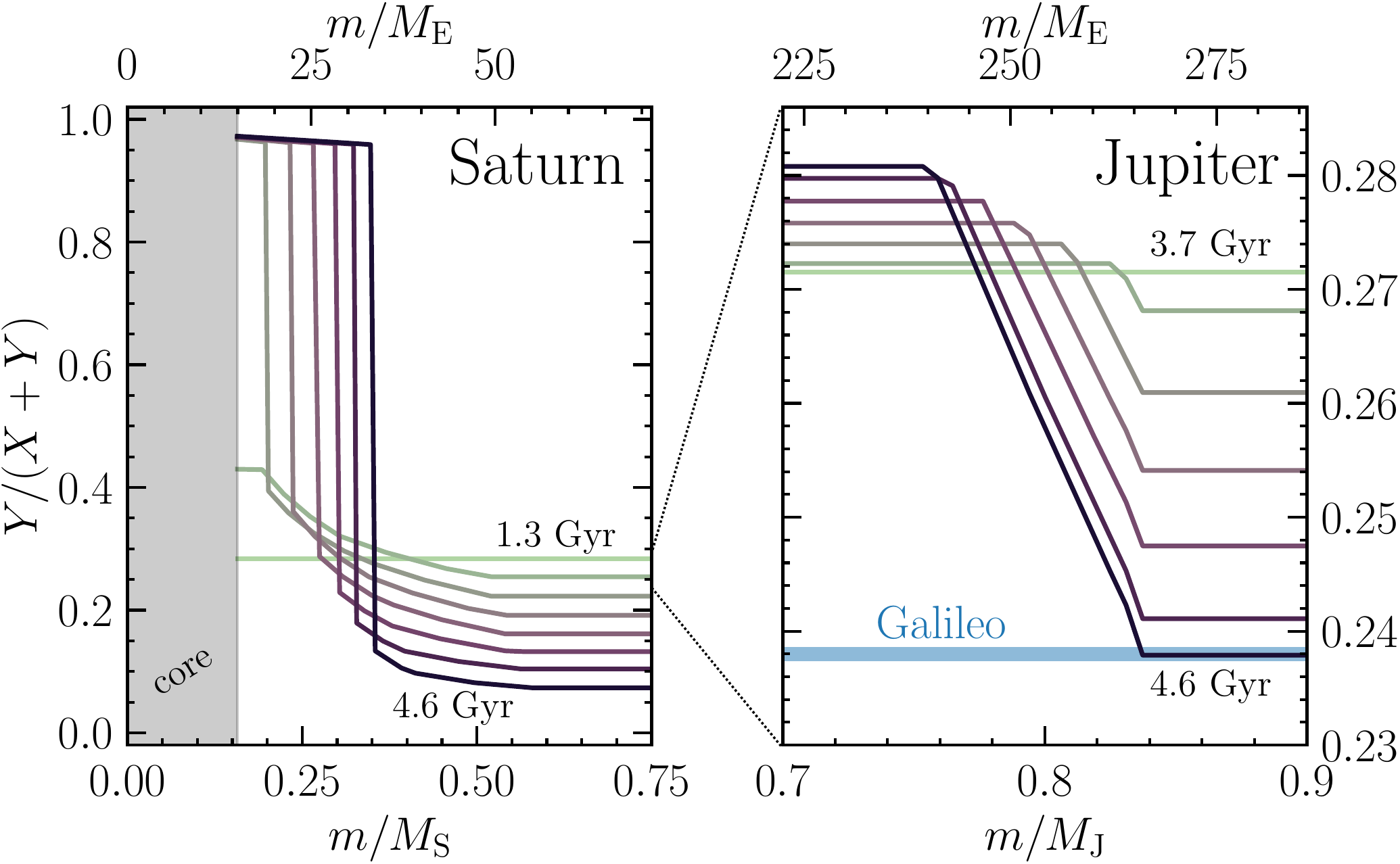}
    \caption{Profiles of the helium mass fraction within Saturn (left) and Jupiter (right) as a function of mass shell and age in evolutionary models \citep{Mankovich20a} coupled to a modified version of the \cite{Schottler18} H/He phase diagram designed to reproduce the Galileo $Y_{\rm atm}$ measurement in Jupiter (\citealt{vonzahn98}; blue line).  The top x-axis is in Earth masses, $M_{\rm E}$. 
    Note the very different scales for the two planets.
    \label{fig.sat_jup_he_MF20}
    }
\end{figure}

A narrow range of fully consistent solutions was indeed found.  These solutions require a modest temperature shift of 500 K hotter He phase separation onset, little to no superadiabacity in either Jupiter or Saturn, and an upward revision Saturn's Bond albedo to be $\sim$0.5, in line with a published upward revision to Jupiter's value from \cas\ fly-by data \citep{Li18}.  Of particular note in these models is that He phase separation proceeds differently in Saturn than in Jupiter (see Fig.~\ref{fig.sat_jup_he_MF20}).  In Jupiter, the He phase separation region takes up very little mass within the planet, and the vast reservoir of hot liquid metallic hydrogen allows He-rich droplets to re-dissolve, slightly enriching the metallic region in He overall.

However, in Saturn He phase separation begins far earlier in the planet's evolution, and the separation region affects a far larger region of the planet in mass-shell-space.  Essentially all of the liquid metallic hydrogen region reaches He immiscibility due to Saturn's much lower pressure at the bottom of the H/He envelope compared to Jupiter.  This quickly leads to He-rich droplets ``pooling up" atop the core, which in the \citet{Mankovich20a} models was treated as a distinct water/rock core. These models suggest that He phase separation is not merely a perturbation on the planet's overall structure (as in Jupiter) but radically alters the mass distribution and cooling history of Saturn.  It is therefore clear that in order to better understand the evolution and internal structure of Saturn improved knowledge of the H/He phase diagram is required, and most importantly, a measurement of the He abundance in Saturn's atmosphere.

\subsubsection{Uncertainty in the initial conditions and relation to formation}
Another type of Saturn evolution model includes composition gradients in the deep interior \cite[e.g.,][]{LC2013,Vazan2016}. Such interiors are non-adiabatic which could be a result of a primordial composition gradient due to the formation process itself \citep{Lozovsky2017,Helled17,Valletta2020} and/or because of the erosion of a central core and immiscibility effects (e.g., of helium in metallic hydrogen). 
Composition gradients can inhibit convection and influence the heat transport inside Saturn. 
If the composition gradient is steep and the luminosity is not very large, overturning convection is suppressed and mixing and homogenization are prevented. In that case the heat transport would be dominated by conduction/radiation or a less efficient type of convection (layered/fingering convection). 

\begin{figure}[ht]
	\centering
	\includegraphics[width=0.48\textwidth]{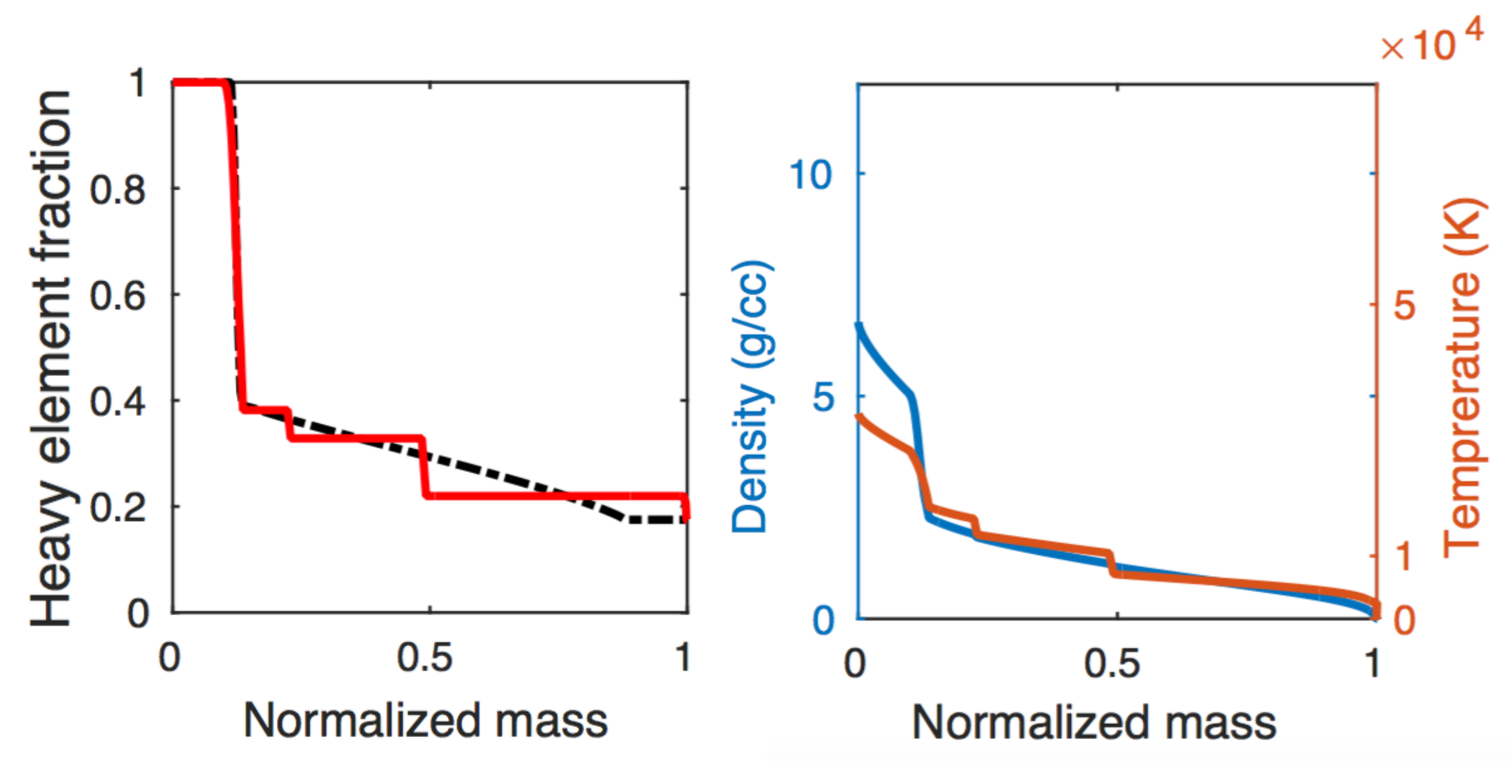}
    \caption{An example of a not fully-adiabatic model of Saturn. Left: The primordial (dashed-black) and current-state (solid-red) heavy-element distribution in Saturn.  
    Right: The density (blue) and temperature (red) profiles for the current-state internal structure.  (adapted from \citet{Vazan2016}.)
\label{satnon}}
\end{figure}

If giant planets indeed form with primordial composition gradients, their long-term evolution and current-state internal structure are affected. Using a planet evolution code that accounts for the effect of the heavy-elements on the planet self-consistently, also including helium rain (see \citealt{Vazan2016} for details) it was found that the innermost regions of Saturn are stable against large-scale convection. As a result, the heat cannot be transported efficiently in these regions, while the outer envelope is convective throughout the entire evolution. The increasing temperature gradient between the innermost non-adiabatic and non-convective region and the outer convective region leads to a small convective penetration inward during the long-term evolution, leading to a moderate heavy-element enrichment of the outer regions as time progresses. 
An example for such non-adiabatic models is shown in Figure \ref{satnon}, which shows an example of a Saturn model with primordial composition gradients and its inferred current-state internal structure.  Of note in this model that metal enrichment in the outer half of the H/He envelope is not representative of the entire envelope.

To summarize, Saturn's interior might be well represented by distinct layers, but instead, is expected to have more complex structure which includes composition gradients. These two different possible structures are sketched in Fig.~\ref{sketch}.


\subsection{Current understanding of structure in light of magnetic field data} \label{magmag}
In principle, Saturn's magnetic field provides information about the internal structure that is not accessible by any other means at present because the dynamo process responsible for the field implies aspects of the convection and layering that are not directly obtainable from gravity or even seismology. Importantly, the magnetic field provides insight to the dynamics (i.e. velocity field) of the conducting interior. 
This section is a brief summary; please see the extensive review by Cao et al. elsewhere in the volume.  

Many of the basic questions about the field are lacking a good explanation at present, including the following:
\begin{enumerate}
    \item  Why is the field so small? The dipole field is twenty times smaller than Jupiter's. 
    \item  Why is the field (and its higher harmonics) so spin-axisymmetric? 
    \item Why is the field spectrum so different from other planets? 
    \item How is the existence of this field reconciled with claimed structures inferred from gravity and ring seismology?
\end{enumerate}

There are a number of complications that make addressing these questions difficult.  
In the ``classical'' view of the Saturnian field there is a layer, possibly quite thick, of metallic hydrogen.  
The expected convective motions can easily provide the magnetic Reynolds number $\sim$100 needed for field regeneration \citep{Stevenson03}. In this simple view, Saturn should not differ that much from Earth or the ``classical" view of Jupiter. Importantly, there is no static stability in this picture since vertical motions of substantial radial extent are needed to sustain a dynamo \citep{Busse75}. 


Neither Jupiter nor Saturn are this simple, and the differences between the two planets might lie in these complications. 
First, there is adequate electrical conductivity in dense, hot molecular hydrogen for a dynamo effect or at least for magnetohydrodynamic (MHD) effects. 
This is an old idea \citep{Smoluchowski75}, and is supported by the Juno results \citep{Bolton17}. 
It permits dynamo generation out to larger radii than previously supposed, perhaps relevant to explaining the power observed in higher harmonics and perhaps permitting dynamo action in more than one layer rather than in a single thick uniform convecting shell.

Furthermore, helium differentiation promotes the presence of a layer that inhibits large vertical motions.  Jupiter has strong evidence for helium rain from Galileo probe results, while Saturn has possibly contradictory evidence even though Saturn should exhibit the larger effect because of its presumed lower internal entropy. Saturn's heat flow permits but does not demand a helium rain-out explanation. Still, a layer of helium rain-out is expected in Saturn and is likely more substantial than in Jupiter, favoring a more complex dynamo activity, with field generation both above and below a layer that may inhibit large scale vertical motions.

Cassini gravity (like Juno gravity) clearly detected deep-seated differential rotation. 
The visible winds on Saturn extend to a much deeper level (but perhaps to a similar pressure and temperature) as those on Jupiter. 
It is not known whether this leads to a different MHD or dynamo effect for the two planets.

Lastly, some aspect of their formation has led to compositional gradients in the heavy elements with consequences for convective structure. The inference from ring seismology \citep{Mankovich21} and satellite evolution \citep{Lainey20} of static stability (g-modes) and a dilute core provides additional support for a more complicated layered structure or layered dynamo within Saturn. Since molecular hydrogen may support dynamo action for temperature exceeding a few thousand degrees, especially if differential rotation is substantial and aids field generation, there may be no difficulty in creating compositional gradients out to 0.6 of Saturn's radius, as suggested by \citet{Mankovich21}.

Returning then to the questions posed:
\begin{enumerate}
    \item The difference in field strength between Jupiter and Saturn is not yet explained but hints at the failure of simple scaling laws, perhaps because structure matters and it is not sufficient to characterize a planet by mass density, rotation, heat flow, conductivity and size\citep{Stevenson03}.
    \item The spin-axisymmetry is certainly promoted by the presence of differential rotation \citep{Stevenson80} but it's not yet clear whether this is a sufficient explanation; it could be that the dynamo operates differently under the conditions needed to filter the non-symmetric terms. Moreover, there is no fundamental difficulty with a dynamo that is externally symmetric and the non-axisymmetry could be present in regions or at length scales that are not currently observable.  Either the helium rain region or the overlying differential rotation in molecular hydrogen may be involved.
    \item The non-power-law character to the power spectrum might suggest a layered dynamo with no single ``radius of dynamo generation."
    \item More complicated dynamo models are suggested for both Jupiter and Saturn, meaning that a single shell picture could be insufficient. 
\end{enumerate}



\subsection{Conclusions and Future Work}
Often, Saturn is viewed simply as a smaller version of Jupiter. While both planets have many similarities, they also have fundamental differences and each planet is clearly unique.  The \emph{Cassini} Grand Finale orbits provided an unprecedented data set on Saturn's gravitational field that has yielded strong new constraints on the planet's interior structure.  Studies of ring features linked to normal mode oscillations in the planet's interior are still yielding new results that will continue to provide novel, complementary constraints on the interior structure.

At some point, no more new ring features will be found, and our ``complete" \emph{Cassini} data set will be the precise gravity field and interior normal mode constraints.  At that point no new additional data may be forthcoming for quite some time.  New advances on those longer time horizons will likely have to come from theory.  The EOS of hydrogen, and helium, and their mixtures, from ab initio calculations, vetted by high-pressure experiments, will progress.  Experimental constraints on the H/He phase diagram can give us greater confidence in the effects of phase separation on the planet.

Measurements of the elemental abundances in giant planet atmospheres is critical to further constraining their formation, evolution, and internal structure. While it is still unclear whether such a measurements represent the planetary bulk, the enrichment of specific elements and the measurement of the noble gases can be used to put limits on enrichment mechanisms tied to planetary origin \cite[e.g.,][and references therein]{Mousis2016}. 
As a result, it is desirable to send a entry probe into Saturn's atmosphere, comparable to Jupiter's Galileo probe, and measure the abundance of noble gases (in particular helium and neon) and other key elements and isotopes.  Perhaps the most important measurement for structure models would be the helium abundance, as a definitive, precise measurements in Saturn, to go along with Jupiter, strongly constrains the role of He phase separation in the current structure of Saturn and the role of He sedimentation in the cooling history of the planet \citep{Mankovich20a}.

A complete picture of the planet today, that melds seismology and gravity field data, including differential rotation, and the role of He phase separation, is still a work in progress.  It is significant that seismology has had such an important role is understanding the deepest regions of the planet, where the gravity field tells us less than we would hope.  Clearly, the \cas\ revolution for Saturn's interior is still yet unfinished, but we should be energized by the significant advances that \cas\ has delivered.


\end{document}